\documentclass{mn2e}

\usepackage{amsfonts}
\usepackage{amsmath}
\usepackage{graphicx}

\title[Environmental Dependence of Morphology \& Colour]
      {Galaxy Zoo: Disentangling the Environmental Dependence of Morphology and Colour\thanks{This publication has been made possible by the participation of more than 100,000 volunteers in the Galaxy Zoo project.  Their contributions are individually acknowledged at \texttt{http://www.galaxyzoo.org/Volunteers.aspx}.}}
\author[R. A. Skibba, S. P. Bamford, R. C. Nichol, C. J. Lintott, et al.]
 {Ramin A. Skibba$^1$\thanks{E-mail: skibba@mpia.de}, Steven P. Bamford$^{2,3}$, Robert C. Nichol$^2$, 
Chris J. Lintott$^4$, \newauthor
Dan Andreescu$^5$, Edward M. Edmondson$^2$, Phil Murray$^6$, M. Jordan Raddick$^7$, \newauthor
Kevin Schawinski$^8$, An\v{z}e Slosar$^9$, Alexander S. Szalay$^7$, Daniel Thomas$^2$,\newauthor
Jan Vandenberg$^7$\\
  $^{1}$Max-Planck-Institute for Astronomy, K\"{o}nigstuhl 17,
	D-69117 Heidelberg, Germany\\
  $^{2}$Institute of Cosmology and Gravitation, University of Portsmouth, Mercantile House,
        Hampshire Terrace, Portsmouth, PO1 2EG, UK\\
  $^{3}$Centre for Astronomy and Particle Theory, University of Nottingham,
        University Park, Nottingham, NG7 2RD, UK\\
  $^{4}$Astrophysics, University of Oxford, Denys Wilkinson Building,
        Keble Road, Oxford, OX1 3RH, UK\\
  $^{5}$LinkLab, 4506 Graystone Ave., Bronx, NY 10471, USA\\
  $^{6}$Fingerprint Digital Media, 9 Victoria Close, Newtownards, Co. Down,
        Northern Ireland, BT23 7GY, UK\\
  $^{7}$Department of Physics and Astronomy, The Johns Hopkins University,
        Homewood Campus, Baltimore, MD 21218, USA\\
  $^{8}$Yale Center for Astronomy and Astrophysics, Yale University,
        P.O. Box 208121, New Haven, CT 06520, USA\\
  $^{9}$Berkeley Center for Cosmo. Physics, Lawrence Berkeley National Lab. \& Physics Dept.,
        Univ. of California, Berkeley CA 94720, USA
}

\newcounter{appfig}

\begin{document}

\pagerange{\pageref{firstpage}--\pageref{lastpage}}

\maketitle
\label{firstpage}

\begin{abstract}
We analyze the environmental dependence of galaxy morphology and colour
with two-point clustering statistics,
using data from the Galaxy Zoo, the largest sample of visually classified 
morphologies yet compiled,
extracted from the Sloan Digital Sky Survey.
We present two-point correlation functions of spiral and early-type galaxies,
and we quantify the correlation between morphology and environment with
marked correlation functions.
These yield clear and precise environmental trends across a wide range of scales,
analogous to similar measurements with galaxy colours,
indicating that the Galaxy Zoo classifications themselves are very precise.
We measure morphology marked correlation functions at fixed colour and find 
that they are relatively weak, with the only residual correlation being that
of red galaxies at small scales, indicating a morphology gradient within haloes
for red galaxies.
At fixed morphology, we find that the environmental dependence of colour
remains strong, and these correlations remain for fixed morphology \textit{and}
luminosity.
An implication of this is that much of the morphology--density relation
is due to the relation between colour and density.
Our results also have implications for galaxy evolution: 
the morphological transformation of galaxies is usually accompanied by
a colour transformation, but not necessarily vice versa.
A spiral galaxy may move onto the red sequence of the colour-magnitude diagram
without quickly becoming an early-type.
We analyze the significant population of red spiral galaxies, and present
evidence that they tend to be located in moderately dense environments
and are often satellite galaxies in the outskirts of haloes.
Finally, we combine our results to argue that central and satellite galaxies
tend to follow different evolutionary paths.
\end{abstract}


\section{Introduction}

The galaxies in the local universe appear to be classifiable into two broad types:
late-type galaxies, with spiral arms, disk-dominated morphologies, ongoing star 
formation, and blue optical colours;
and early-type galaxies, with elliptical or bulge-dominated morphologies, old stellar
populations, and red colours.
Many studies have observed a strong correlation between morphology and environment
(\textit{i.e.}, the morphology--density relation), with spiral galaxies
tending to be located in low-density environments and elliptical galaxies
in more dense environments (e.g., Dressler 1980, Postman \& Geller 1984,
Norberg et al. 2002, Goto et al. 2003, Blanton et al. 2005, Wolf et al. 2007, Ball et al. 2008).
Other bimodal galaxy properties have also been found to be strongly correlated 
with the environment: blue galaxies with significant star formation activity
tend to reside in underdense environments, while galaxies with red colours and
low star formation rates tend to reside in overdense environments 
(e.g., Lewis et al. 2002, G\'{o}mez et al. 2003, Kauffmann et al. 2004, Zehavi et al. 2005, 
Weinmann et al. 2006, Sheth et al. 2006, Tinker et al. 2008, Bamford et al. 2008, Skibba \& Sheth 2009).

In standard $\Lambda$CDM cosmological models, cold dark matter haloes form from 
the gravitational collapse of dark matter around peaks in the initial density field. 
Haloes assemble 
hierarchically, such that smaller haloes merge to form larger and more massive
haloes in dense environments (Mo \& White 1996, Sheth \& Tormen 2002).
According to the current paradigm of galaxy formation, galaxies form within 
haloes, due to the cooling of hot gas.
Halos and galaxies evolve simultaneously, and the evolution of a galaxy is 
affected by its host halo.   If the halo is accreted by a larger halo, the
galaxy will be affected by it as well:
for example, the galaxy's diffuse hot gas reservoir may be stripped,
removing its fuel for future star formation (e.g., 
Larson, Tinsley \& Caldwell 1980; Balogh, Navarro \& Morris 2000;
Weinmann et al. 2006; van den Bosch et al. 2008b).  
Galaxies may also experience major mergers, which transform late-type
galaxies into early-type galaxies with a central bulge component (e.g., Driver et al. 2006, Drory \& Fisher 2007).
Mergers drive gas towards the center, where it can trigger a burst of star formation
and fuel the central black hole, the feedback from which can heat the remaining gas and 
eventually quench star formation (e.g., Mihos \& Hernquist 1996, Wild et al. 2007, 
Pasquali et al. 2008, Schawinski et al. 2009a).

In this general picture of galaxy formation and evolution, it is expected
that galaxies' morphological and star formation properties are differently
correlated with the environment.
The purpose of this paper is to shed light on some of these galaxy evolution processes
by disentangling the environmental correlations of morphology and colour.
We use mark statistics of galaxy clustering, which are sensitive to environmental
correlations (e.g., Skibba et al. 2006),
with catalogues of galaxies in the Sloan Digital Sky Survey (SDSS; York et al. 2000)
to explore the respective roles of morphology and colour in galaxy evolution.
We characterise morphologies using classifications from the Galaxy
Zoo project (as described in Section~\ref{GZmorph}).


This paper is complementary to the work by Bamford et al. (2009),
who studied the environmental dependence of morphology and colour with similar catalogues,
using projected galaxy density within an adaptive radius and distance to 
the centre of the nearest galaxy group to characterize the environment.
Bamford et al. used these environmental indicators to obtain information about 
the fractions of galaxy morphologies and colours as an explicit function of 
environment. 
However, for their local density estimator, a single density corresponds to a range of scales,
while the use of the distance to the nearest group implicitly assumes group-specific environmental mechanisms, in addition to the assumptions of the group-finding algorithm.
Our mark clustering statistics yield complementary information, quantifying environmental trends as an explicit function of scale.
This paper and Bamford et al. provide very different statistical descriptions of the same dataset.
In addition, Bamford et al. explored the environmental dependence of morphology and colour
at fixed stellar mass, while we explore the environmental dependence of morphology at fixed colour, and vice versa.
Both papers also shed light on the substantial populations of `unconventional' galaxies:
red spirals and blue early-types.
Finally, both the mark correlation functions presented here and the correlations 
with group mass and luminosity in Bamford et al. can be compared to halo occupation models
and semi-analytic models in a straightforward manner.

This paper is organized as follows.
In the next section, we describe the Galaxy Zoo data and our volume-limited catalogues.
We introduce mark clustering statistics, and in particular, the marked correlation function,
in Section~\ref{markstats}.
In Section~\ref{morphmarks}, we present marked correlation functions using Galaxy Zoo
morphology likelihoods as marks, which show the environmental dependence of spiral
and early-type morphologies, and of merging and interacting galaxies. 
Next, we disentangle the environmental dependence of morphology and colour
in Section~\ref{morphcolour}, by analyzing morphology-environment correlations
at fixed colour and colour-environment correlations at fixed morphology.
We focus on the populations of red spiral galaxies and blue early-type galaxies in
Section~\ref{redspirals}, and our results allow us to explain their roles in galaxy evolution.
Finally, we summarize our results in Section~\ref{discuss}.

\section{Data}\label{data}

\subsection{Basic Catalog Properties}

The catalogues used in this paper are constructed from the SDSS Data Release 6 (DR6; Adelman-McCarthy et al. 2008).
We consider only the galaxies in the Main Galaxy Sample, extended objects with
$r_\mathrm{petro}<17.77$ (Strauss et al. 2002).
We also only use galaxies with spectroscopic redshifts.

This raises the issue of fiber collisions, where the thickness of the spectroscopic fibers means that galaxies closer than 55" on the sky will be missing spectra and redshifts, and hence, absolute magnitudes.
This fiber-collision constraint is partly alleviated by the fact that neighbouring plates have overlap regions, but it still results in $\sim7\%$ of targeted galaxies not having a measured redshift.
Fiber collisions weakly affect correlation functions at scales of $s\sim100\,\mathrm{kpc}/h$ in redshift-space, and mark correlation functions as well, if galaxy spectra are required for accurate marks (Skibba et al. 2006).
However, we measure only \textit{projected} marked correlation functions, in which
any effects of fiber collisions cancel out (see Eq.~\ref{markedwp} in Section~\ref{markstats}).
In addition, the fiber assignments were based solely on target positions, and in
cases where multiple targets could only have a single fiber assigned, the target
selected to be observed was chosen randomly--hence independently of galaxy properties.
For the projected correlation functions $w_p(r_p)$, we do not observe any downturn at small scales, 
which one might expect, due to the higher fraction of missing galaxies.
In addition, we have performed tests with similar catalogs with different minimum redshifts, and have also compared some measurements to those of Skibba \& Sheth (2008), in which fiber-collided galaxies were included and were given the redshift of their nearest neighbour; in both cases, the correlation functions were consistent at small scales.
We conclude that the effects of fiber collisions are negligible for both the projected correlation functions and marked projected correlation functions.

We use Petrosian magnitudes for the galaxy luminosities and colours.
Absolute magnitudes are determined using \texttt{kcorrect v4\_1\_4} (Blanton \& Roweis 2007)
and are $k$-corrected and corrected for passive evolution to $z=0.1$.
We restrict our analysis to volume-limited samples, described in Section~\ref{volltd}.

Throughout this paper we assume a spatially flat cosmology with 
$\Omega_m=0.3$ and $\Omega_\Lambda=1-\Omega_m$. 
We write the Hubble constant as $H_0=100h$~km~s$^{-1}$~Mpc$^{-1}$.

\subsection{Galaxy Zoo Morphologies}\label{GZmorph}

Although galaxy colours are directly measurable, their morphologies are not.
A variety of galaxy properties, such as those that quantify the radial light
profile of a galaxy, have been used as proxies for galaxy morphologies
(e.g., Blanton et al. 2005, van den Bosch et al. 2008a).
Automated methods for classifying galaxies using more structural information
have also been recently developed
(e.g., Conselice 2006, Park et al. 2007). 
However, 
none of these techniques can yet provide a direct equivalent of traditional visual morphology.
Studies such as Lahav et al. (1996) and Ball et al. (2004) have attempted to reproduce visual
morphologies with artificial neural networks.
However, their classifications rely heavily on luminosity and colour information, whereas
true visual morphology is determined without direct reference to these quantities.
The Galaxy Zoo project was born out of a need for reliable, visual morphologies
for a large sample of SDSS galaxies.
Our approach was to enlist the public's help to visually classify the morphologies of
nearly one million galaxies 
in SDSS DR6 (Adelman-McCarthy et al. 2008).
Further details of the Galaxy Zoo project, including its motivation, design, and
the initial stages of the data reduction, are given in Lintott et al. (2008).
We use morphological type likelihoods estimated from the
Galaxy Zoo classifications (Bamford et al. 2009) throughout this paper.

The morphologies utilized here are derived from classifications by over 80,000
members of the international public as part of the Galaxy Zoo project.  This project is
described in detail by Lintott et al. (2008).
Briefly, each galaxy received many, independent morphological classifications, each
by a different user.
The four possible classifications were labeled as `elliptical', `spiral', `don't know',
and `merger'.
The `elliptical' class, in addition to containing galaxies with elliptical morphology,
also contains the majority of S0 galaxies (Bamford et al. 2009).
The merger class mainly comprises interacting pairs, generally with tidal features.
The spiral classification was subdivided into `clockwise', `anti-clockwise', and
`edge-on/unsure', referring to the direction and orientation of the spiral arms.

The median number of classifications per object is 34, with 98 per cent of our sample
having at least 20 classifications each.  These classifications were processed into
raw `likelihoods' for a galaxy being of a given morphological type, directly from the
proportion of classifications for each type.  

The raw Galaxy Zoo dataset is affected by two redshift-dependent biases.
The first is the usual selection bias, due to the apparent magnitude and 
angular size selection limits of the parent SDSS Main Galaxy Sample.  In 
this paper we avoid the issue of selection bias by considering only 
volume-limited samples.  The second effect is classification bias: 
identical galaxies are more likely to be classified as early-type with 
increasing redshift, due to signal-to-noise and resolution effects.  
Bamford et al. (2009) have quantified the classification bias as a 
function of galaxy luminosity, size and redshift, which they then use to 
statistically correct the Galaxy Zoo data.  Note that absolutely no 
reference was made to the environment of the galaxies at any stage of 
this classification bias correction.  Throughout this paper we utilize 
the de-biased type likelihoods, which we denote $P_\mathrm{sp}$ and 
$P_\mathrm{el}$ for spiral and early-type likelihoods, respectively.

\subsection{Volume-Limited Catalogs}\label{volltd}

We perform our analysis on a volume-limited large-scale structure sample.
Our catalogue has limits $-23.5< {^{0.1}M_r}<-19.5$, $0.017<z<0.082$,
consisting of 99924 galaxies, with mean number density $\bar n_{\mathrm {gal}}=0.011\,(h^{-1}\mathrm{Mpc})^3$.
The superscript of 0.1 refers to the fact that all absolute magnitudes have
been $k$-corrected and evolution-corrected to $z=0.1$; henceforth, we omit
the superscript.
This luminosity threshold approximately corresponds to $M_r<M^\ast+1$,
where $M^\ast$ is the Schechter function break in the 
$r$-band luminosity function (Blanton et al. 2003).
It also corresponds to an approximate halo mass threshold of 
$5.8\times10^{11}\,h^{-1}\,M_\odot$ 
(Skibba \& Sheth 2009). 
We have performed the same analysis with a more luminous volume-limited catalogue,
with $M_r<-20.5$, and have obtained results qualitatively similar to those
presented in Sections~\ref{morphmarks}-\ref{redspirals}.

There are several `false pairs' of objects in the SDSS---large objects 
with more than one ID associated with them.  This results in extra close
pairs being counted.  We avoid these objects by calculating their angular separation, and then if the distance is less than either of their Petrosian radii, we exclude the smaller `false' object.
There were only a handful of such objects in our volume-limited catalogues.

Most of our analysis is focused on spiral and elliptical galaxies.
To avoid `contamination' by other galaxies, we excluded galaxies with
high merger or `don't know' likelihoods, with
$P_\mathrm{mg}$ or $P_\mathrm{dk}>15$ per cent or with 
$P_\mathrm{sp}+P_\mathrm{el}<75$ per cent in the measurements, except when stated otherwise.
These are conservative cuts, and they reduce the sample size of the catalogues by $~\sim10$ per cent.

For the measured correlation functions and jack-knife errors, which 
require random catalogues and jack-knife sub-catalogues, we use the 
hierarchical pixel scheme SDSSPix
\footnote{\texttt{http://lahmu.phyast.pitt.edu/$\sim$scranton/SDSSPix}},
which characterizes the survey geometry, including edges and holes 
from missing fields and areas near bright stars.
This same scheme has been used for other clustering analyses (Scranton et al. 2005, 
Hansen et al. 2007) and for lensing analyses (Sheldon et al. 2007).
We use the Scranton et al. \texttt{jack\_random\_polygon} with window files from 
SDSS DR5 (Adelman-McCarthy et al. 2007), which reduces the sample size by an additional 
$\sim20$ per cent, to a final size of 72135 galaxies. 
For the random catalogues, we used at least ten times as many random
points as in the data, and for each of the error calculations, 
we used thirty jack-knife sub-catalogues.
The jack-knife errors of our marked projected correlation functions tend to be larger than the
Poisson errors, even at small scales.  We note that Norberg et al. (2008) has 
recently shown that the jack-knife method does not recover the scale dependence of errors
of the unmarked correlation function, often overestimating the errors at small scales, and the results are sensitive to the number of sub-catalogues into which the data is split.
Although our uncertainty estimates are important, our primary results are not particularly
sensitive to the precise value of the errors.

\section{Mark Clustering Statistics and Environmental Correlations}\label{markstats}

We characterize galaxies by their properties, or `marks', such as their 
luminosity, colour, morphological type, stellar mass, star formation rate, etc.
In most galaxy clustering analyses, a galaxy catalogue is cut into subsamples 
based on the mark, and the two-point clustering in the subsample is studied by 
treating each galaxy in it equally (e.g., Norberg et al. 2002, 
Madgwick et al. 2003, Zehavi et al. 2005, Li et al. 2006, Tinker et al. 2008).
These studies have shown that galaxy properties are correlated with the 
environment, such that elliptical, luminous, and redder galaxies tend to be 
more strongly clustered than spiral, fainter, and bluer galaxies.
For example, Norberg et al. (2002) divided their catalogue by luminosity
and spectral type into subsamples, measured their correlation functions,
and found that luminous and quiescent galaxies have larger clustering strengths.

However, the galaxy marks in these studies are used to define the subsamples for the 
analyses, but are not considered further.
This procedure is not ideal because the choice of critical threshold for 
dividing galaxy catalogues is somewhat arbitrary, and because throwing away the 
actual value of the mark represents a loss of information.
In the current era of large galaxy surveys, one can now measure not only galaxy clustering as a function of their properties, but the spatial correlations of the galaxy properties themselves.
We do this with `marked statistics', in which we weight each galaxy by a particular mark,
rather than simply count galaxies as `ones' and `zeroes'.

Marked clustering statistics have recently been applied to astrophysical datasets by
Beisbart \& Kerscher (2000), Gottl\"{o}ber et al. (2002), and Faltenbacher et al. (2002).
Marked statistics are well-suited for identifying and quantifying correlations 
between galaxy properties and their environments (Sheth, Connolly \& Skibba 2005).
They relate traditional unmarked galaxy clustering to the clustering in which each galaxy is weighted by a particular property.
Marked statistics are straightforward to measure and interpret: 
if the weighted and unweighted clustering are significantly different at a particular scale,
then the galaxy mark is correlated (or anti-correlated) with the environment at that scale,
and the degree to which they are different quantifies the strength of the correlation.
In addition, issues that plague traditional clustering measurements, such as incompleteness and complicated survey geometry, do not significantly affect measurements of marked statistics, as these effects cancel out to some extent, since the weighted and unweighted measurements are usually similarly affected.
Finally, the halo model framework has been used to interpret the correlations of luminosity and colour marks in terms of the correlation between halo mass and environment (Skibba et al. 2006, Skibba \& Sheth 2009).
We focus on morphological marks here, and although they could be similarly analyzed with the halo model, perhaps by building on these luminosity and colour mark models, that is beyond the scope of this paper.

There are a variety of marked statistics (Beisbart \& Kerscher 2000, Sheth 2005), but 
the easiest to measure and interpret is the marked two-point correlation function.
The marked correlation function is simply 
\begin{equation}
  M(r) \,\equiv\, \frac{1+W(r)}{1+\xi(r)},
 \label{markedXi}
\end{equation}
where $\xi(r)$ is the two-point correlation function, 
the sum over galaxy pairs separated by $r$, in which all galaxies are `weighted' by unity.
$W(r)$ is the same sum over galaxy pairs separated by $r$, but 
now each member of the pair is weighted by the ratio of its mark to 
the mean mark of all the galaxies in the catalogue 
(e.g., Stoyan \& Stoyan 1994).
That is, for a given separation $r$, $\xi(r)$ receives a count of 1 for each galaxy pair,
and $W(r)$ receives a count of $W_i W_j$ for $W(r)$.
The fact that the marked statistic $M(r)$ can usually be accurately estimated
by the simple pair count ratio $WW/DD$ 
(where $DD$ are the counts of data-data pairs and $WW$ are the weighted counts), 
without requiring a random galaxy catalogue, 
implies that the marked correlation function is less sensitive than the
unmarked correlation function to the effects of the survey edges (Sheth, Connolly \& Skibba 2005).
In effect, the denominator in Eqn.~\ref{markedXi} divides out the contribution to the 
weighted correlation function which comes from the spatial 
contribution of the points, leaving only the contribution from the 
fluctuations of the marks.  That is, the mark correlation function
measures the clustering of the marks themselves, in $r$-scale environments.

In practice, in order to obviate issues involving redshift distortions,
we use the projected two-point correlation function
\begin{equation}
  w_p(r_p)\,=\, \int {\mathrm d}r\,\xi({r_p},\pi)\, 
            = \,2\, \int_{r_p}^\infty \,{\mathrm d}r\,
                         \frac{r\,\xi(r)}{\sqrt{r^2-{r_p}^2}},
\end{equation}
where $r=\sqrt{{r_p}^2+\pi^2}$, $r_p$ and $\pi$ are the galaxy 
separations perpendicular and parallel to the line of sight, and 
we integrate up to line-of-sight separations of $\pi =40\,\mathrm{Mpc}/h$.
We estimate $\xi({r_p},\pi)$ using the Landy \& Szalay (1993) estimator
\begin{equation}
  \xi({r_p},\pi) \,=\, \frac{DD-2DR+RR}{RR},
\end{equation}
where $DD$, $DR$, and $RR$ are the normalized counts of data-data, 
data-random, and random-random pairs at each separation bin.
Similarly, the weighted projected correlation function is measured by
integrating along the line-of-sight the analogous weighted statistic 
\begin{equation}
  W(r_p,\pi) \,=\, \frac{WW-2WR+RR}{RR},
\end{equation}
where $W$ refers to a galaxy weighted by some property (for example, 
$P_\mathrm{sp}$ or $P_\mathrm{el}$), and $R$ now refers to an
object in the catalog of random points, weighted by a mark chosen randomly from its distribution.
We then define the marked projected correlation function:
\begin{equation}
  M_p(r_p)\,=\, \frac{1\,+\,W_p(r_p)/r_p}{1\,+\,w_p(r_p)/r_p} \, ,
 \label{markedwp}
\end{equation}
which makes $M_p(r_p)\approx M(r)$ on scales larger than a few Mpc.
The projected correlation functions $w_p$ and $W_p(r_p)$ are normalized
by $r_p$, so as to be made unitless.
On large scales both the real-space and projected marked correlation
functions (Eqns~\ref{markedXi} and \ref{markedwp}) will approach unity, 
because at increasing scale the correlation functions $\xi(r)$ and $W(r)$ 
(or $w_p(r_p)$ and $W_p(r_p)$) become very small as the universe appears nearly homogeneous.
The simple ratio of the weighted to the unweighted correlation function $W(r)/\xi(r)$
(or $W_p(r_p)/w_p(r_p)$) approaches unity similarly, provided that 
there are sufficient number statistics and the catalogue's volume is sufficiently large.

\section{Environmental Dependence of Galaxy Morphology}\label{morphmarks}

We explore the environmental dependence of galaxy morphology in Galaxy Zoo
in this and the next section by measuring marked projected correlation functions 
of galaxies in volume-limited catalogues.

When measuring mark clustering statistics, it is important to carefully choose
marks that are appropriate for one's analysis.
For example, traditional morphological types of galaxies are discrete marks (e.g., E, S0, Sp, and Irr in Norberg et al. 2002; \textit{cf.}, Wolf et al. 2007, Poggianti et al. 2008).
Continuous marks can be more useful for analyses of mark correlations, but like discrete marks,
they must be measurable with sufficient precision such that contamination by measurement error is not a serious concern.
A variety of continuous marks have been used to attempt to characterize
the morphological or structural properties of galaxies, such as
Sersic index, surface brightness, $r_{90}/r_{50}$ concentration, 
and their combinations (Blanton et al. 2005, Christlein \& Zabludoff 2005, Conselice 2006, Scarlata et al. 2007, Park et al. 2007, van den Bosch et al. 2008a, Ball et al. 2008, van der Wel 2008).
However, while such structural parameters are no doubt informative,
they provide limited insight concerning the presence of a disk or spiral
arms, and are thus not equivalent to traditional visual morphology;
rather, they are often more strongly related to galaxy luminosity or
colour (see discussion in Bamford et al. 2009).

Here we use the morphological-type \textit{likelihoods} estimated in the Galaxy Zoo (Lintott et al. 2008; see Section~\ref{GZmorph}) as marks.
The likelihoods were estimated using the visual classifications of the Galaxy Zoo, and are
corrected for classification bias (Bamford et al. 2009). 
These probabilities reflect the certainty with which Galaxy Zoo
classifiers were able to assign each galaxy to either of the discrete
``spiral'' or ``elliptical'' classes, rather than a continuous measure of
morphology.  This choice of mark thus encodes our level of knowledge
concerning the morphology of each galaxy.  The Galaxy Zoo morphologies
have been shown to be reliable by comparison with datasets visually
classified by professional astronomers.  Note that we find that
galaxies which are classified as lenticulars by professional
astronomers are predominantly placed in the ``elliptical'' Galaxy Zoo
class with high likelihood, and we thus henceforth refer to this class
as ``early-type''.  As estimates of visual morphology in the standard
and generally accepted sense, we believe the Galaxy Zoo
classifications to be more suitable for our analysis than the various automatic
morphology proxies that have been proposed.  The Galaxy Zoo
type-likelihoods, rather than discrete types or combinations of structural properties,
are a reliable 
choice of mark for exploring the dependence of galaxy morphology on environment.

\begin{figure}
 \includegraphics[width=\hsize]{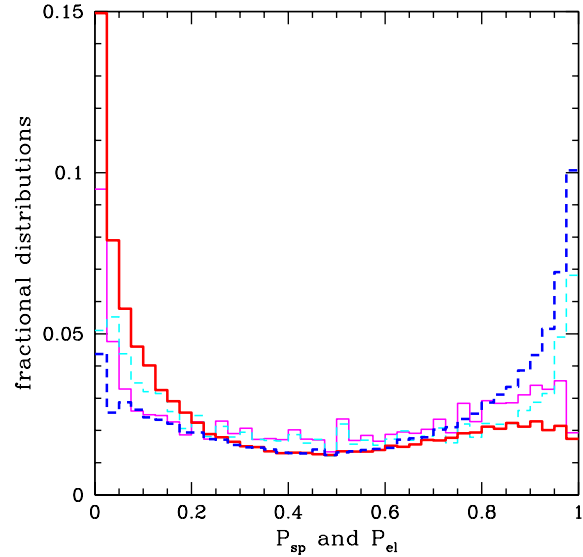} 
 \caption{Distributions of morphology marks for $M_r<-19.5$.
          Thick red (solid) and blue (dashed) histograms show the distributions of 
          $P_\mathrm{el}$ and $P_\mathrm{sp}$, respectively, and 
          thin magenta (solid) and cyan (dashed) histograms show the distributions of their 
          respective uncorrected raw probabilities, prior to the corrections for 
          classification bias.}
 \label{probdists}
\end{figure}

Mark clustering statistics are sensitive to the distributions of the marks.
We begin by presenting morphology mark distributions in Figure~\ref{probdists}, for the $M_r<-19.5$ volume-limited catalogue.
The morphology likelihoods $P_\mathrm{sp}$ and $P_\mathrm{el}$ are required to be between 0 and 1, and they both are somewhat peaked at the minimum and maximum values,
which simply reflects the fact that there is usually good agreement between
Galaxy Zoo classifiers on the classification of each object.
The figure also clearly shows that the distributions are smoother for the likelihoods
corrected for classification bias, discussed in Section~\ref{GZmorph}.
In what follows, we only use the de-biased type-likelihoods for our analysis.

\begin{figure}
 \includegraphics[width=\hsize]{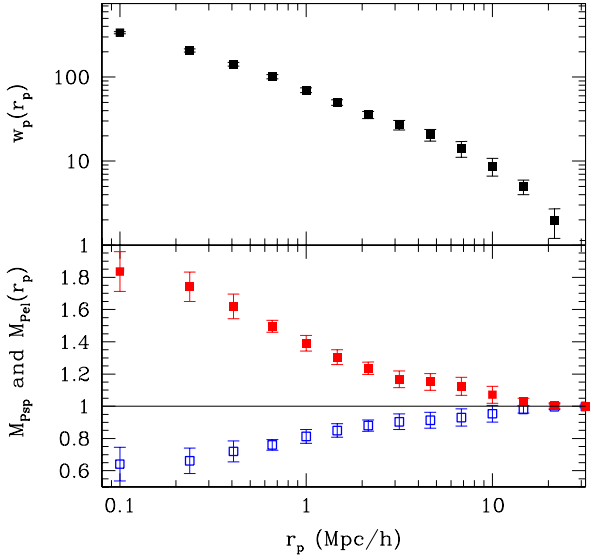} 
 \caption{Projected two-point correlation function (upper panel) and mark correlation 
          functions (lower panel) for $M_r<-19.5$.  Red (filled) and blue (open) points show the 
          measurements for the $P_\mathrm{el}$ and $P_\mathrm{sp}$ marks,
          with jack-knife error bars.
         }
 \label{MCF195}
\end{figure}

Now we present our morphology mark correlation functions for $M_r<-19.5$ in Figure~\ref{MCF195}.
At projected separations of $r_p<10\,\mathrm{Mpc}/h$, the mark correlations of
$P_\mathrm{el}$ are significantly above unity, indicating that it is correlated with
the environment at these scales; that is, early-types tend to be located in denser environments.
In contrast, the mark correlations of $P_\mathrm{sp}$ are significantly below unity, indicating that it is anti-correlated with the environment---that is, spirals tend to be located in underdense environments.
The scale-dependence of the mark correlations is remarkably smooth, monotonically increasing or decreasing towards unity at large scales.  The smoothness of these measurements, with no spikes or bumps, is partly due to the large sample size, but it also indicates that the marks have very little noise.
This is impressive in itself.
Morphological classifications are often very uncertain, but the likelihoods from a large number of classifiers, with robust corrections for `classification bias', appear to be sufficiently precise to yield clear environmental correlations.

The mark correlation functions are a useful tool to show the environmental dependence
of morphology, both of pairs of galaxies within a single dark matter halo (on small 
scales, $r_p\leq2\,\mathrm{Mpc}/h$), and of pairs in separate haloes (on large scales).
In contrast, most other studies of morphology and environment have focused on the surface density within one particular scale (e.g., $1\,\mathrm{Mpc}/h$) or have used the projected distance or surface density out to the $N^\mathrm{th}$ nearest neighbour, which mixes different scales together.
In Figure~\ref{MCF195}, a transition appears to occur in the $P_\mathrm{el}$ marked
correlation function at $r_p\sim2\,\mathrm{Mpc}/h$, at the scale of the transition 
from the `one-halo' to the `two-halo' term.
At scales much larger than the size of a typical halo, the scale dependence of marked
statistics is simply related to the shape of the linear theory power spectrum (Sheth 2005).
The large-scale marked correlation functions are related to morphology
dependent galaxy bias.


The stellar masses of galaxies are correlated with their morphologies and with
the environment (e.g., Balogh et al. 2001, Bamford et al. 2009).
It is therefore possible that our observed correlation between morphology and environment is
simply due to the environmental dependence of stellar mass.
In order to test this, we use stellar masses estimated from luminosity and
colour, $r_\mathrm{petro}$ and $(u-r)_\mathrm{model}$ (Baldry et al. 2006), and 
measure the morphology mark correlation functions at fixed stellar mass.
The stellar masses have uncertainties of $\sim0.2$ dex.  We use stellar
mass intervals of 0.3 dex: $10.2\leq\mathrm{log}(M_\ast/M_\odot)<10.5$,
$10.5\leq\mathrm{log}(M_\ast/M_\odot)<10.8$, and $10.8\leq\mathrm{log}(M_\ast/M_\odot)<11.1$.
These subsamples contain 26037, 22581, and 9791 galaxies, respectively.

\begin{figure}
 \includegraphics[width=\hsize]{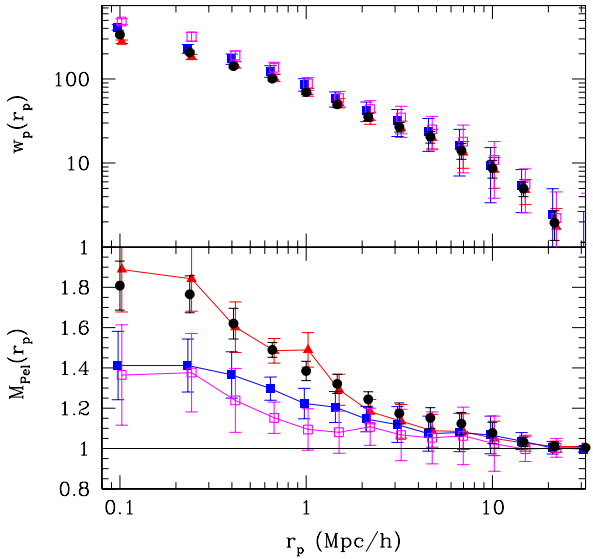} 
 \caption{Projected two-point correlation function (upper panel) and $P_\mathrm{el}$ mark
          correlation functions (lower panel) for $M_r<-19.5$, for
          stellar mass intervals: $10.2\leq\mathrm{log}(M_\ast/M_\odot)<10.5$ (triangles),
          $10.5\leq\mathrm{log}(M_\ast/M_\odot)<10.8$ (filled squares), and
          $10.8\leq\mathrm{log}(M_\ast/M_\odot)<11.1$ (open squares).
          Black circles show result for full catalogue (repeated from Fig.~\ref{MCF195}).}
 \label{MCF195mstar}
\end{figure}

The resulting clustering measurements are shown in Figure~\ref{MCF195mstar}.
The environmental correlations remain very strong, and for the lowest mass
interval, the strength of the correlation is not reduced at all.
This is similar to the result obtained by Bamford et al. (2009), 
that the environmental dependence of the early-type fraction is strong
at fixed stellar mass, especially for low-mass galaxies.
Even for the most massive galaxies, we find that the environmental correlation is
significant out to a few $\mathrm{Mpc}/h$.
The figure shows the result for the $P_\mathrm{el}$ mark; the result is
approximately the same for $P_\mathrm{sp}$: the correlation between
morphology and environment remains strong even at fixed mass.
Therefore, a galaxy of a given stellar mass will be more likely to be an
early-type (and less likely to be a spiral) in a dense environment compared
to a galaxy of the same mass in an underdense environment.
Importantly, our stellar masses are estimated from luminosity and colour,
both of which are correlated with morphology and the environment 
(e.g., Blanton et al. 2005, Park et al. 2007).
In addition, the environmental dependence of stellar mass, as characterized by 
mark correlations, can be entirely described by the combined environmental 
dependence of luminosity and colour (Skibba \& Sheth, in prep.).
%
It has long been known that 
the morphologies and colours of galaxies are closely related (e.g., Hubble 1936, de Vaucouleurs 1961, Strateva et al. 2001), 
and we attempt to disentangle the environmental correlations
of these properties in Section~\ref{morphcolour}.

For all of these mark correlation measurements, 
in order to focus on the population of galaxies most likely to be spirals or early-types, 
we have excluded galaxies with $P_\mathrm{mg}$ or $P_\mathrm{dk}>15$ per cent or with $P_\mathrm{sp}+P_\mathrm{el}<75$ per cent.
Consequently, for nearly all galaxies in the catalogue, the sum of $P_\mathrm{sp}$ and 
$P_\mathrm{el}$ is nearly unity (\textit{i.e.}, $1-P_\mathrm{sp}$ is almost
equivalent to $P_\mathrm{el}$).
If these cuts were not made, however, the mark correlations using the $1-P_\mathrm{sp}$
mark is slightly weaker than those of $P_\mathrm{el}$, indicating that
$P_\mathrm{mg}$, or even $P_\mathrm{dk}$, is weakly correlated with the environment.

\begin{figure}
 \includegraphics[width=\hsize]{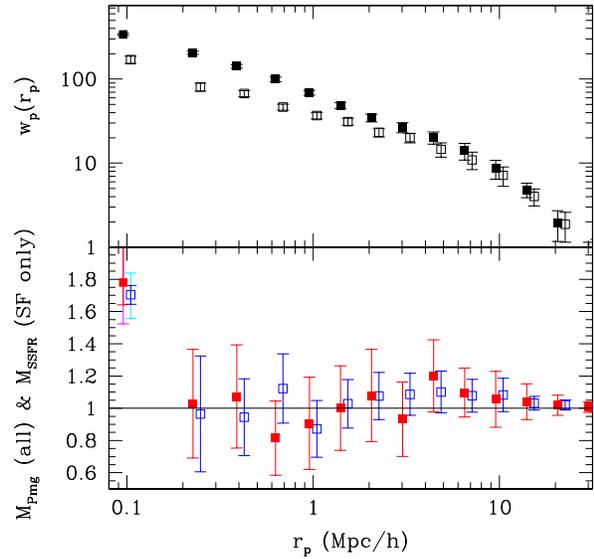} 
 \caption{Projected two-point correlation function and $P_\mathrm{mg}$ mark correlation
          function for all galaxies in the $M_r<-19.5$ catalogue (filled square points).
          For comparison, measurement from Skibba, Wake \& Sheth (in prep.) of 
          $w_p(r_p)$ and $SFR/M_\ast$ $M(r_p)$ of star-forming galaxies is also shown (open square points).
          For $M(r_p)$ at $100\,\mathrm{kpc}/h$, error bars show $1$ and $2~\sigma$
          of the distributions of measurements from jack-knife sampling.} 
 \label{Pmergplot}
\end{figure}

We show the measurement of the $P_\mathrm{mg}$ mark correlation function
in Figure~\ref{Pmergplot}, now with no cuts on the morphological type-likelihoods.
$P_\mathrm{mg}$ quantifies the likelihood with which a galaxy is classified as 
undergoing an interaction or merger, usually with indications of tidal distortions or tails;
heavily distorted objects may be more likely to be classified with a high $P_\mathrm{dk}$ (`don't know').
The $P_\mathrm{mg}$ likelihoods are generally quite accurate, such that
every galaxy in the Galaxy Zoo with $P_\mathrm{mg}\geq0.6$ has been found
to show clear evidence of merging (Darg et al. 2009a).
The probability that a galaxy is observed to be undergoing a merger or interaction 
is approximately uncorrelated with the environment except at scales of 
$r_p\sim100\,\mathrm{kpc}/h$, where it spikes upwards at 
$M(r_p)=1.78_{-0.25}^{+0.42}$. (This point has a wide bin width 
of galaxy separations, $1.75<\mathrm{log}(r_p/\mathrm{kpc}h^{-1})<2.25$, 
though we obtain a similar spike when we repeat the measurement using a 
narrower bin and when we vary only the lower or upper limit of the bin.)
We have also measured the marked correlation function on even smaller scales,
and $M(r_p)$ appears to rise to $\approx3$ at $r_p=50\,\mathrm{kpc}/h$ and
$\approx30$ at $r_p=25\,\mathrm{kpc}/h$, though such measurements are very uncertain. 

Interestingly, this is very similar to the mark correlation function of the Brinchmann et al. (2004) specific star formation rate $SFR/M_\ast$ mark measured by Skibba, Wake \& Sheth (in prep.), for a similar $M_r<-19.5$ volume-limited catalogue (open points in Fig.~\ref{Pmergplot}).
Their measurement is for highly star-forming galaxies \textit{only}, with strong
emission lines ($H_\alpha$, $H_\beta$, [O III], [N II], with AGN excluded),
which is why there is no large-scale trend.
The extinction-corrected $SFR(H_\alpha)$ mark correlation function 
also displays similar behaviour.
This result is consistent with the picture that mergers and interactions between galaxies 
can drive gas into their central regions, triggering bursts of star formation.
In a thorough investigation of merging and interacting galaxies in Galaxy Zoo, Darg et al.
(2009a) have similarly found that the specific SFRs are about twice as high in star-forming
merging galaxies than in star-forming control galaxies.
Our result is also consistent with Robaina et al. (2009), who find a similar 
small-scale enhancement of star formation at $z\sim0.6$, also using
visually classified morphologies, with SFRs estimated from UV and IR luminosity.

The correlation between $P_\mathrm{mg}$ and the environment is not unexpected:
mergers and interactions are relatively common in groups and clusters, and rare
in underdense environments.
It is interesting that the mark correlations are significant out to
$r_p\approx170\,\mathrm{kpc}/h$ (the maximum separation for the bin centered on $r_p=100\,\mathrm{kpc}/h$), which is near the virial radius of the minimum
halo mass associated with the luminosity threshold of the galaxy catalog,
estimated from halo occupation modeling (see Skibba \& Sheth 2009).
This suggests that much of the signal is due to galaxy pairs 
interacting and influencing each other in the central regions of dark matter haloes 
with masses above this threshold.
However, this scale should be interpreted with caution, because the effects
of galaxy mergers and interactions on morphologies and star formation 
are expected to be important on much smaller scales ($r<40\,\mathrm{kpc}$).
Our measurements do not imply that these effects persist out to larger scales: 
for example, two galaxies $i$ and $j$ with a separation $170\,\mathrm{kpc}/h$ 
and a relatively large mark product $P_{\mathrm{mg},i} P_{\mathrm{mg},j}$ 
might not be merging with each other.  Rather, they may both be located in a 
small-scale environment in which the occurrence of mergers is relatively likely. 
%
The scale of the upturn in the mark correlation function has 
a simple physical explanation: mergers tend to occupy slightly denser 
environments than average, often in galaxy groups (McIntosh et al. 2008, 
Darg et al. 2009b).  Galaxy pairs in groups are often found at projected 
separations of $r_p<200\,\mathrm{kpc}/h$, and some of them may have recently 
experienced a merger or interaction.  Galaxy mergers also occur in 
clusters, where galaxies often have larger projected separations, but the 
fraction of galaxies that are merging is relatively low.

We also note that, our result in Figure~\ref{Pmergplot}, while suggestive, 
does not imply a \textit{necessary} connection
between the processes of merging and star formation:
the merging/interacting and starbursting galaxy pairs are not necessarily the same,
and even if they are, we cannot by these methods establish a causal connection. 
Because the mark distributions are so different and because of the 
uncertainties involved, we cannot draw stronger conclusions at this time.

Note that the Galaxy Zoo merger/interaction likelihoods have other potential uses as well.
For example, they could be used to estimate merger rates or fractions,
as has been done with other morphological classifications 
(e.g., de Propris et al. 2007, Lotz et al. 2008).
Galaxy close pair
counts have been used extensively for this purpose, but $P_\mathrm{mg}$ \textit{weighted}
pair counts could also contribute to such studies, or could be used as tests for other
methods.  We can define the following as the fraction of galaxies within a distance $r_f$
that show signs of merging (e.g., distortions or tidal tails that indicate a significant
gravitational interaction):
\begin{equation}
 f_\mathrm{merg}(r<r_f) \,=\,
   \frac{4\pi {\bar n}_\mathrm{gal}\, \langle P_\mathrm{mg}\rangle^2\, \int_0^{r_f}\mathrm{d}r\,r^2\,(1+W_{P_\mathrm{mg}}(r))}
   {4\pi {\bar n}_\mathrm{gal} \int_0^{r_f}\mathrm{d}r\,r^2\,(1+\xi(r))} ,
 \label{fmerg}
\end{equation}
where the denominator is the close pair fraction, (see Masjedi et al. 2006, Bell et al. 2006)
and $W_{P_\mathrm{mg}}(r)$ is the \textit{real-space} $P_\mathrm{mg}$ 
weighted correlation function, normalized by the mean mark $\langle P_\mathrm{mg}\rangle$ 
(which is approximately 0.027 for the catalogue used for Fig.~\ref{Pmergplot}) squared.
If the real-space weighted and unweighted correlations can be reliably 
estimated at small scales, then this fraction can be estimated as well.
However, our measured $P_\mathrm{mg}$ weighted correlation function is very
uncertain at such small scales, 
and robustly estimating the value of $f_\mathrm{merg}$ is beyond the scope of this paper.

Finally, Slosar et al. (2008) have detected a chiral correlation
function in the spins of spiral galaxies at small galaxy separations.
Using tidal torque theory, they interpret this result as correlated angular momenta.
However, it is also plausible that spiral arms could be enhanced or even
induced by galaxy-galaxy interactions in a manner that is consistent with
their signal.  This could be tested with a mark cross-correlation function,
to determine whether spiral chirality and disturbed morphology are correlated
at small scales. 
In a related issue, Jimenez et al. (2009) recently found a correlation between 
past star formation activity and the degree of coherence of spin direction, 
especially for galaxies that formed most of their stars in the past.
The fact that nearby spiral galaxies may have similar star formation histories and 
aligned spins is interpreted as the result of the large-scale environment influencing 
both halo spins and star formation activity.


\section{Disentangling Morphology and Colour}\label{morphcolour}

The previous section showed the environmental dependence of galaxy morphology 
using mark correlation functions with the Galaxy Zoo. 
It is well-known, however, that morphology and colour are strongly correlated (e.g., Blanton et al. 2005, Ball et al. 2008, van der Wel 2008),
and that galaxy colour is also correlated with the environment (e.g., Zehavi et al. 2005, Berlind et al. 2006, Cooper et al. 2008, Coil et al. 2008, Tinker et al. 2008, Wang et al. 2007, Skibba \& Sheth 2009, Bamford et al. 2009;
\textit{cf.}, Weinmann et al. 2006 and Skibba 2009 for the different correlations of central and satellite galaxies).   
As galaxies evolve, their star formation properties and structural properties are expected to transform differently (van den Bosch et al. 2008a); therefore, if we can disentangle the dependence of morphology and colour on the environment, we can shed some light on aspects of galaxy evolution.
The main purpose of this section is to separate the environmental dependence of morphology and colour, and we do this by analyzing morphology mark correlation functions at fixed colour, and colour mark correlation functions at fixed morphology.

\subsection{Environmental Dependence of Morphology at Fixed Colour}\label{fixedcolour}

\begin{figure}
 \includegraphics[width=\hsize]{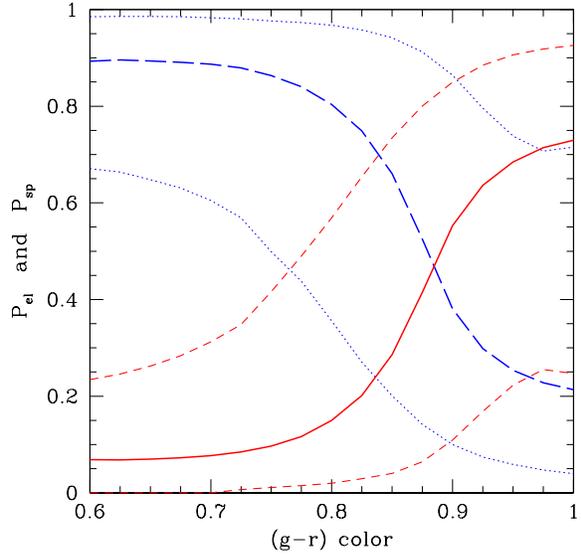} 
 \caption{Median correlation between $P_\mathrm{el}$ likelihood and $g-r$ colour (red solid curve), and 
          $P_\mathrm{sp}$ and $g-r$ (blue long-dashed curve) in the $M_r<-19.5$ volume-limited catalogue.
          Thin lines show the 16 and 84 percentiles.}
 \label{Pellgminr}
\end{figure}

We begin by explicitly showing the correlation between the Galaxy Zoo likelihoods $P_\mathrm{el}$ and $P_\mathrm{sp}$, and $g-r$ colour, in Figure~\ref{Pellgminr}.
The correlation is strongest for likelihoods between 0.3 and 0.7, and in the colour range $0.8<g-r<0.95$.
Note also that there is significant scatter across the whole range in colour and morphology likelihood, but especially at the red end of the plot.
We have also measured the morphology \textit{fractions} as a function of colour (see Bamford et al. 2009),
which also indicate that, while most blue galaxies are spirals, a significant fraction of red galaxies are not early-types.

\begin{figure}
 \includegraphics[width=\hsize]{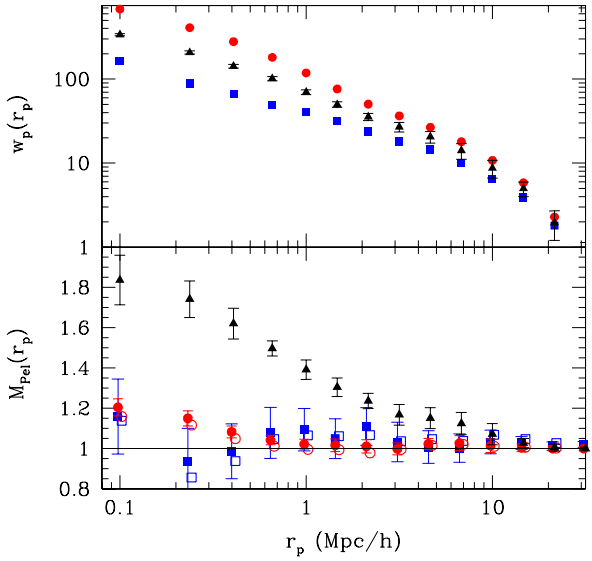} 
 \caption{Projected correlation function (upper panel) and $P_\mathrm{el}$ mark
          correlation function 
          (lower panel) of all galaxies in the $M_r<-19.5$ catalogue (black triangles, from Fig.~\ref{MCF195})
          compared to that of red and blue galaxies (red circle points and blue square points, respectively).
          `Red' and `blue' galaxy subcatalogues are determined using the colour-magnitude cut in
          Eqn~\ref{colourcut} (filled points) or using the colour likelihood
          thresholds $P_\mathrm{red}>0.8$ and $P_\mathrm{blu}>0.8$ (open points).
         }
 \label{morphMCFredblu}
\end{figure}

In Figure~\ref{morphMCFredblu}, we show the projected clustering of all galaxies,
and of red and blue galaxies separately, using the following colour-magnitude cut
(used by Blanton \& Berlind 2007, Skibba \& Sheth 2009):
\begin{equation}
  {(g-r)}_\mathrm{cut} \,=\, 0.8 \,-\, 0.03\,({M_r}+20) .
  \label{colourcut}
\end{equation}
Note that this is different than the $u-r$ versus $M_r$ colour-magnitude cut used by
Bamford et al. (2009); we do not expect our results to be sensitive to the 
exact choice of the cut.

The upper panel of the figure simply shows that red galaxies are more strongly
clustered than blue galaxies.
The lower panel shows the environmental dependence of morphology at fixed colour.
In particular, we compare the $P_\mathrm{el}$ mark correlation function for all galaxies 
(from Figure~\ref{MCF195}) to the $P_\mathrm{el}$ mark correlation functions
for red and blue galaxies (circle and square points).
The morphology mark signal is considerably weakened at fixed colour.

Therefore, \textit{the correlations between morphology and environment 
are extremely weak 
at fixed colour}:
blue galaxies have no morphology-environment correlation, and for red galaxies,
only a weak residual correlation at small scales remains.
This implies that much of the correlation between morphology and environment
is explained by that of colour and environment, and the remaining environmental 
dependence occurs only on scales smaller than the diameter of a typical dark matter halo.

Colour gradients or `segregation' within haloes, defined with respect to the halo radii,
have been observed in group and cluster catalogues 
(Hansen et al. 2007, van den Bosch et al. 2008b), and they 
appear to be approximately independent of halo mass.
(Note that colour gradients are much stronger than luminosity gradients 
(Hansen et al. 2007); since galaxy stellar masses are well-described by 
their luminosities and colours (Bell et al. 2003), 
this suggests that, at least for some subpopulations of galaxies, the
environmental dependence of their stellar populations is such that
the observed stellar mass segregation within groups (van den Bosch et
al. 2008b) manifests itself observationally merely as colour segregation.)
Such gradients as a function of halo-centric or group-centric distance result 
in slightly stronger mark correlations, but only at small scales (Skibba 2009).
The fact that the residual correlation in Figure~\ref{morphMCFredblu} occurs only at scales of $r_p<500\,\mathrm{kpc}/h$, 
that is, predominantly within dark matter haloes, 
suggests that in haloes of given mass, red early-types are more strongly clustered than the
whole red galaxy population. 
It is possible that this correlation is related to stellar mass segregation in haloes, 
because although Bamford et al. (2009) observe a similar small-scale 
morphology-environment correlation at fixed stellar mass, it is relatively weak.
However, if the morphology gradient were mostly due to stellar mass segregation, then at fixed stellar mass, most of the morphology-environment correlation should disappear at small scales, but this is clearly not the case (see lower panel of Figure~\ref{MCF195mstar}).

In any case, we interpret the residual mark correlation in Figure~\ref{morphMCFredblu} 
such that, haloes may host both
red early-types (red galaxies with high $P_\mathrm{el}$) and red spirals 
(red galaxies with low $P_\mathrm{el}$), but the latter tend 
to be located further from the halo center and are more likely to be `satellite' galaxies.
The galaxy pairs at small separations consist mostly of central-satellite pairs, and therefore red satellites closer to a red central galaxy in a given halo are more likely to be early-types than red satellites further away.
If there were a significant halo mass dependence of morphology at fixed colour, then the marked
correlations would have to be statistically significant (i.e., inconsistent with 
$M(r_p)=1$) on larger scales---scales at which the clustering statistics are dominated 
by the contribution from pairs in separate haloes---but we do not observe such a 
correlation in Figure~\ref{morphMCFredblu}.
In the halo model, massive dark matter haloes are more strongly clustered than less massive haloes 
(e.g., Mo \& White 1996); when large-scale galaxy clustering depends on a galaxy property, 
indicated by significant large-scale marked correlations, this implies that the property, 
at least for central galaxies, is correlated with halo mass (Sheth 2005, Skibba et al. 2006, 
Skibba \& Sheth 2009), but this is not the case here, for morphology at fixed colour.

It follows from this argument that galaxy morphologies depend more on colour and 
halo-centric distance than on halo mass,
so at fixed (red) colours, we expect a morphology gradient within haloes.
For example, we expect that $P_\mathrm{morph}(r/r_\mathrm{vir}|g-r,M_\mathrm{halo})$ is not constant for galaxies with red $g-r$,
which is a prediction that can be tested with satellite galaxies in group catalogues.
We have also observed a somewhat similar $P_\mathrm{el}$ correlation for
red galaxies in a more luminous catalogue, and hence for galaxies in more massive haloes,
so it is possible that the strength of this gradient is approximately independent of halo mass 
(for haloes with $M_\mathrm{halo}>6\times10^{11}\,M_\odot/h$),
but our results are not conclusive on this point.
If this is the case, then it would imply that the transformation of red spiral
satellite galaxies into red lenticular and elliptical galaxies occurs similarly
in all environments.  
On the other hand, if the transformation mechanisms that change these galaxies' 
morphologies, such as ram-pressure stripping, galaxy merging and tidal `harassment', depend 
on halo mass, then they may be more (or less) efficient in more massive haloes, 
resulting in a different morphology gradient in haloes of different mass.

We have also performed similar measurements with a stricter colour-magnitude cut, using
colour \textit{likelihoods}, analogous to the morphology likelihoods.
In particular, we use the bimodal $g-r$ colour distribution as a function of
$r$-band luminosity: the red sequence and its rms, the blue sequence and its rms, and 
the blue fraction, as a function of luminosity (Skibba \& Sheth 2009).
With this information, we estimate each galaxy's $P_\mathrm{red}$ and $P_\mathrm{blu}$
(normalized to total unity) based on its position on the $(g-r)\,-\,M_r$ colour-magnitude diagram.
When we use the stricter cuts of $P_\mathrm{red}>0.8$ and $P_\mathrm{blu}>0.8$
(open points in Fig.~\ref{morphMCFredblu}), the result is the same:
at fixed colour, there is almost no morphology-environment correlation.
This complements the result from Bamford et al. (2009), who found that at fixed stellar mass (determined from $u-r$ colour and $r$-band luminosity), there is only a weak residual correlation between early-type fraction and projected density and almost no correlation between early-type fraction and projected distance to the nearest galaxy group (their Figures 11 and 14).

Note that the distributions of morphology likelihoods are different for red and blue
galaxies.  Since mark clustering statistics are sensitive to the distributions of the marks,
it is possible that the weakening of the environmental correlations in 
Figure~\ref{morphMCFredblu} is simply due to differences between the distributions.
However, we show in the appendix 
that, even after accounting for the different mark distributions, the correlation
between morphology and environment at fixed colour is still relatively weak.

\subsection{Environmental Dependence of Colour at Fixed Morphology}\label{fixedmorph}

\begin{figure}
 \includegraphics[width=\hsize]{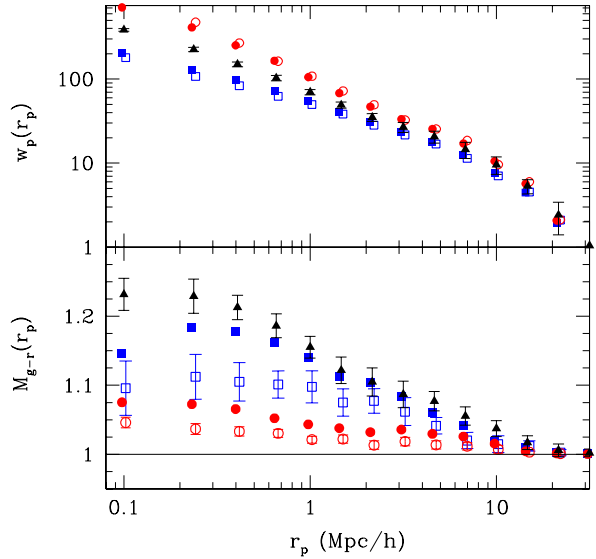} 
 \caption{Projected correlation function (upper panel) and $g-r$ colour mark correlation function
          (lower panel) of all galaxies in the $M_r<-19.5$ catalogue (triangle points, from Skibba \& Sheth 2009) 
          compared to that of early-type and spiral galaxies (red circle points and blue square points, respectively).
          The spiral and early-type subcatalogues are determined using the greater morphology
          likelihood as the distinguishing criterion (filled points)
          or using the likelihood threshold of $P>0.8$ (open points). }
 \label{colourMCFellspi}
\end{figure}

We now perform a similar analysis of the environmental dependence of galaxy colour at fixed morphology. 
In Figure~\ref{colourMCFellspi}, we show the projected clustering of all galaxies, and of early-type and spiral galaxies separately.
The upper panel simply shows that early-type galaxies are more strongly clustered than spirals.
The lower panel shows the environmental dependence of colour at fixed morphology,
analogous to Figure~\ref{morphMCFredblu} in Section~\ref{fixedcolour}.
In particular, we compare the $g-r$ colour mark correlation function of all galaxies (from Skibba \& Sheth 2009) to that of early-types and spirals.
The lower values of $M(r_p)$ here compared to Figure~\ref{morphMCFredblu} are simply
due to the different distributions of colour and morphology likelihoods.
The galaxy colours are always within a factor of 1.5 of the mean colour, while the
morphology likelihoods vary by more than an order of magnitude.  Therefore,
the mark product of the colours of a galaxy pair in a dense environment may be not be that much larger than the mark product of a pair in an underdense environment, while 
the difference between the pairs' morphology likelihood marks may be much larger.

If one simply counts spiral galaxies as those with $P_\mathrm{sp}>P_\mathrm{el}$,
and early-types as the rest, then the resulting colour mark correlations are somewhat 
weakened (solid square points), but are still very significant out to large scales 
($r_p\sim10\,\mathrm{Mpc}/h$).
If one applies stricter thresholds, with $P_\mathrm{sp}>0.8$ for spirals and $P_\mathrm{el}>0.8$ for early-types, then the mark correlations are weakened slightly further (open square points), but they remain significant out to large scales.
This result is robust, and it is not affected by the different colour distributions of spirals and early-types (Appendix~\ref{markdistapp}).
The fact that we see strong mark correlations out to large-scale separations
in Figure~\ref{colourMCFellspi}, in contrast with Figure~\ref{morphMCFredblu}, 
implies that at fixed morphology, galaxy colours are correlated with the environment: 
at a given morphology likelihood, galaxies tend to be redder in more massive systems 
(e.g., clusters) than in less massive systems (e.g., groups).

\begin{figure}
  \includegraphics[width=\hsize]{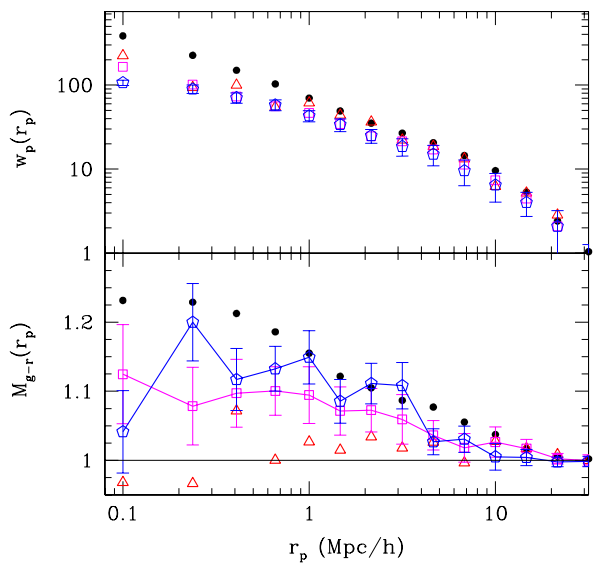} 
  \caption{Projected correlation functions and $g-r$ mark correlation functions,
           as in Figure~\ref{colourMCFellspi}, but for spiral galaxies ($P_\mathrm{sp}>0.8$)
           in three luminosity bins: $-20<M_r<-19.5$ (blue pentagons), $-20.5<M_r<-20$ 
           (magenta squares), and $-21<M_r<-20.5$ (red triangles).
           Error bars are shown only for the fainter two subsamples, and the points are
           connected, for clarity.}
  \label{colourMCFfaintspi}
\end{figure}

We have also measured colour mark correlation functions at fixed morphology \textit{and}
luminosity, and have found significant residual environmental trends.  For example,
in Figure~\ref{colourMCFfaintspi}, we show the $g-r$ mark correlation function 
of spiral galaxies (with $P_\mathrm{sp}>0.8$) in luminosity bins of $-20<M_r<-19.5$, 
$-20.5<M_r<-20$, and $-21<M_r<-20.5$.  The uncertainties are large because the 
subsamples are small, consisting of 11216, 9022, and 5621 galaxies, respectively.
For the two fainter bins, we detect a statistically significant marked signal out
to large scales ($r_p\approx8\,\mathrm{Mpc}/h$), while the measurement of the third
luminosity bin is consistent with no trend.
We have obtained a similar result for early-type galaxies, with significant environmental
correlations for fainter galaxies and a very weak trend for $-21<M_r<-20.5$.
These results are consistent with those of Bamford et al. (2009):
at faint luminosities, the colours of both spirals and early-types are correlated
with the environment, while at bright luminosities, most galaxies of any morphology
are fairly red, and the narrow colour distribution results in weaker or nonexistent colour mark correlations.
This also explains the weak colour-environment correlation observed by Schawinski et al. (2007),
whose sample of early-types is brighter and smaller than ours.

Therefore, \textit{at fixed morphology and luminosity, a relatively strong correlation between colour and environment remains.}
This is in stark contrast to the result in the previous section, that at fixed colour, almost no morphology-environment correlation remains.
The results in Figures~\ref{morphMCFredblu} and \ref{colourMCFellspi} show that 
the morphology-density relation is to a significant extent 
a \textit{colour}-density relation, 
which is consistent the results from other studies (Blanton et al. 2005, Quintero et al. 2006,
Wolf et al. 2007, Bamford et al. 2009). 
Contrary to these studies and the results in this paper,
Park et al. (2007) claim that the key galaxy property is morphology,
and that at fixed morphology and luminosity, galaxy colour is nearly independent
of the environment.  Their disagreement may be due to the fact that
their morphology classification was explicitly based on galaxy colours and colour gradients,
while we, Bamford et al., and Wolf et al. used visual classifications,
and Blanton et al. and Quintero et al. used S\'{e}rsic index and surface brightness as morphological indicators.

It is worth noting that, as pointed out by Quintero et al. (2006), it is possible in principle
that the strong colour-environment correlation and extremely weak morphology-environment
correlation is due to the measurement errors of the marks.
In particular, it is possible that, (i) the errors of the morphology likelihoods are 
much larger than those of the colours, and in an extreme case, so large that they 
wash out some of the environmental correlations;
and (ii) the errors themselves could be correlated with each other and with the environment.
However, firstly, our error analyses show that the uncertainties in the morphology mark
correlation functions are only slightly larger than those of the colour mark correlation
functions.
Secondly, the Galaxy Zoo morphology classifications were done independently of the environment
and were tested with black and white images (Lintott et al. 2008).
The corrections for classification bias (Bamford et al. 2009) also appear to be robust 
for the redshifts to which our volume-limited catalogue is restricted.
We conclude that the errors of the morphologies and colours can have only a 
minimal impact on our correlation function measurements, and they do not affect our results.

The results in this section have implications for galaxy evolution.
They suggest first that a morphology transformation is usually accompanied by a colour transformation: that is, if a galaxy transforms from a spiral to an early-type, its star formation is quenched and it becomes red on a relatively short time-scale.
On the other hand, a colour transformation can occur without a morphological transformation: for example, a spiral galaxy can move onto the red sequence without becoming an early-type.
Such `red spirals' are the subject of Section~\ref{redspirals}.

\subsection{Combined Environmental Dependence of Morphology and Colour}\label{morphcolourcombo}

\begin{figure}
 \includegraphics[width=\hsize]{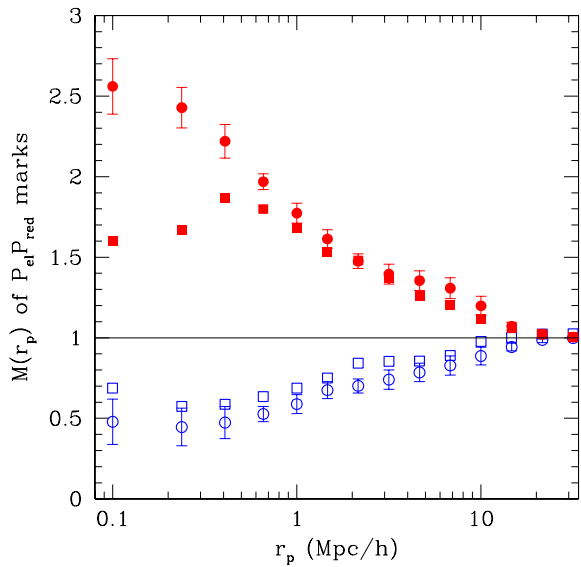} 
 \caption{Morphology-colour mark correlation functions, for the following marks:
          $P_\mathrm{el}P_\mathrm{red}$ (red filled circles), $P_\mathrm{sp}P_\mathrm{red}$
          (red filled squares), $P_\mathrm{el}P_\mathrm{blu}$ (blue open squares),
          $P_\mathrm{sp}P_\mathrm{blu}$ (blue open circles).
          }
 \label{PredPspiusw}
\end{figure}

Finally, we compare the relative strength of the environmental correlations of
colour and morphology in Figure~\ref{PredPspiusw}.
In order to compare more directly to the morphology likelihoods, instead of using the colours of galaxies, we now use their colour likelihoods (described earlier in this section),
estimated from the $(g-r)\,-\,M_r$ colour distribution as a function of luminosity.
Like the colour likelihoods, we now also normalize $P_\mathrm{el}$ and $P_\mathrm{sp}$ 
by $P_\mathrm{el}+P_\mathrm{sp}$ here, so that they also total unity.
The $P_\mathrm{el}$ and $P_\mathrm{red}$ mark distributions, and their effect on
the mark correlations, are discussed in Appendix~\ref{markdistapp}.

Figure~\ref{PredPspiusw} shows the result, where we mark each galaxy by the four morphology-colour 
combinations of $P_\mathrm{el}$ and $P_\mathrm{sp}$, and $P_\mathrm{red}$ and $P_\mathrm{blu}$.
Firstly, by comparing the red and blue circle points to the red and blue points
in Figure~\ref{MCF195}, the environmental correlations for the $P_\mathrm{el} P_\mathrm{red}$
and $P_\mathrm{sp} P_\mathrm{blu}$ marks are clearly magnified, due to the 
combined environmental dependence of both morphology and colour.
That is, red early-types tend to be located in very dense environments, while blue spirals tend to
reside in very underdense environments.
In addition, on scales of $r_p<500\,\mathrm{kpc}/h$, the environmental dependence of the 
red early-types appears to be stronger than the environmental anti-correlations of the
blue spirals: for example, $M_{P_\mathrm{red}P_\mathrm{el}}(r_p)$ (red circles) rises to
$\approx2.5$ at $r_p\approx100\,\mathrm{kpc}/h$ (\textit{i.e.}, the weighted correlation
function is 2.5 times stronger than the unweighted correlation function), while
$M_{P_\mathrm{blu}P_\mathrm{sp}}(r_p)$ drops only to $\approx0.5$ (\textit{i.e.}, the
weighted correlation function is only 2 times weaker than the unweighted correlation function).
This may be an indication that the vast majority of red early-types are central galaxies.

Secondly, for the $P_\mathrm{sp} P_\mathrm{red}$ and $P_\mathrm{el} P_\mathrm{blu}$
marks, one might expect that the mark signal would be weak or nonexistent, as these are 
combining opposite environmental correlations.
Nonetheless, the mark correlations are fairly strong and significant out to large scales, 
and their direction suggests that environmental dependence of colour is stronger than that 
of morphology.
In principle, this could be due to the fact the $P_\mathrm{red}$ distribution is different
from $P_\mathrm{el}$ (which is equivalent to $1-P_\mathrm{sp}$ here).
However, even after accounting for the different distributions (in the appendix),
we still find that galaxies with high $P_\mathrm{red}$ and $P_\mathrm{sp}$
tend to reside in denser environments than average, where the environment
is measured between $100\,\mathrm{kpc}/h<r_p<10\,\mathrm{Mpc}/h$.

There is a downturn in the measurement for the $P_\mathrm{sp} P_\mathrm{red}$ mark
at small scales ($r_p<300\,\mathrm{kpc}/h$),
which may be due to the relatively high satellite fraction of red spirals (discussed in
the next section), resulting in a suppressed central-satellite term, and
also may be partly due to morphology gradients within haloes, if they are stronger
than colour gradients (\textit{i.e.}, if $P_\mathrm{sp}$ decreases more rapidly than 
$P_\mathrm{red}$ increases at small halo-centric radii).
The behaviour of the mark correlation function also demonstrates that the 
morphology likelihoods and colour likelihoods contain independent information,
and that the former is not derivative of the latter.

In any case, the main conclusion from Figure~\ref{PredPspiusw} is that redder, 
more spiral structured galaxies tend to be located in 
denser environments than average, 
and red early-types are even more strongly correlated with the environment.

\section{Red Spiral Galaxies}\label{redspirals} 

While red early-type galaxies and blue late-type galaxies are abundant and well-known,
we now turn to the less understood population of red spiral galaxies.
We simply define red spirals as all those galaxies redder than the colour-magnitude
cut (\ref{colourcut}) and with $P_\mathrm{sp}>0.8$.

We showed in Section~\ref{morphcolour} that at fixed colour the morphology-environment 
correlation is relatively weak, while at fixed morphology some of the 
colour-environment correlation remains.
It follows from this that a morphology transformation of a galaxy usually implies a concomitant colour transformation, so galaxies that transform into early-types while remaining blue should be rare.  In contrast, some galaxies that move onto the red sequence may not undergo a morphology transformation, so there should be a significant population of red spiral galaxies.
For our $M_r<-19.5$ volume-limited catalogue, after excluding galaxies 
with high likelihoods of $P_\mathrm{mg}$ or $P_\mathrm{dk}$,
8443 galaxies out of 90400 are red spirals ($9.3$ per cent),
while only 654 are blue early-types ($0.7$ per cent).
Since red spiral galaxies are relatively common while blue early-types are not, we focus
this section on the red spirals.
Schawinski et al. (2009b) have investigated in detail the blue early-types in the Galaxy Zoo.

It is worth noting that Schiminovich et al. (2007) found, using UV--optical colours
and a dust correction based on 4000\AA break strength and SED fits,
that passive disk-dominated galaxies are relatively rare.
It is therefore possible that a significant fraction of the red spirals
in our sample have some ongoing star formation, but are reddened by dust extinction.
Semi-analytic models also show significant dust extinction among disk-dominated
galaxies, with the dust optical depth having a strong correlation with
gas mass and anti-correlation with scale radius (Fontanot et al. 2008).
Using a combination of IR and UV/optical data, Gallazzi et al. (2008) and Wolf 
et al. (2008) find in the outskirts of the A901/902 supercluster
a significant population of red, yet star-forming, spirals,
with relatively high masses and low Sersic indices. These objects appear red 
due to low star formation rates and their residual star formation being highly 
extincted, and are particularly prevalent at intermediate densities.
We will argue that a large fraction of the red spiral galaxies in the Galaxy Zoo 
are indeed satellite galaxies; their emission line statistics are currently being investigated.

\begin{figure}
 \includegraphics[width=\hsize]{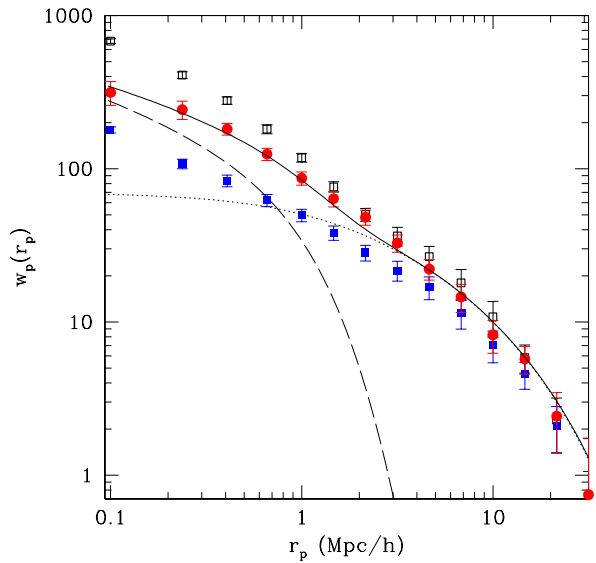} 
 \caption{Projected two-point correlation function of red spiral galaxies (red circles)
          compared to that of all red galaxies (black open squares) and that of 
          all spiral galaxies (blue filled squares) in the $M_r<-19.5$ catalogue.
          A halo occupation model of the correlation function is also shown (solid curve),
          consisting of galaxy pairs in a single halo (dashed curve) and in separate haloes (dotted curve).  The model is consistent with the measurement, with $\chi_\mathrm{d.o.f.}^2=0.5$.
          }
 \label{redspi}
\end{figure}

Figure~\ref{MCF195} showed that spiral galaxies tend to reside in underdense environments, or low-mass dark matter haloes, while Figure~\ref{colourMCFellspi} showed that red galaxies tend to reside in overdense environments, or massive haloes.
We focus now on the galaxies that are both red \textit{and} spiral.
We compare the correlation function of red spiral galaxies to those of 
all red galaxies and all spiral galaxies in Figure~\ref{redspi}.
The clustering of red spirals is consistent with that of red galaxies at large scales.
However, at small scales ($r_p<2\,\mathrm{Mpc}/h$), on the scales of galaxy pairs within the same halo, their clustering signal is significantly depressed.
This result is consistent with Bamford et al. (2009), who found that the
fraction of red spirals, and of red face-on spirals alone, 
increases with projected galaxy density, except in the densest
environments (\textit{i.e.}, in cluster cores), in which the fraction drops.

This suggests that red spirals are often satellite galaxies in the outskirts of groups and clusters, while red early-types tend to be located in the inner regions of these groups.
This is consistent with our interpretation of the weak residual morphology-environment
correlation of red galaxies in Figure~\ref{morphMCFredblu}.

We test this with a halo occupation model of galaxy clustering, (e.g., Zehavi et al. 2005, 
Zheng et al. 2007), in which the halo occupation distribution of central and satellite 
galaxies depends on halo mass and the luminosity threshold.
In Figure~\ref{redspi}, we show the result of such a model, using mean halo occupation 
functions described in Appendix A2 of Skibba \& Sheth (2008).  
%
In simple halo occupation models, for galaxies more luminous than some threshold
($M_r<-19.5$ in this case), corresponding to an approximate halo mass threshold ($M_\mathrm{min}\approx5.8\times10^{11}\,h^{-1}\,M_\odot$),
haloes are occupied by a single central galaxy and $N_\mathrm{sat}$ satellite
galaxies, where $\langle N_\mathrm{sat}|M,L_\mathrm{min}\rangle \approx (M/M_1)^\alpha$,
$M_1\propto M_\mathrm{min}(L_\mathrm{min})$ and $\alpha$ increases with $L_\mathrm{min}$.
In practice, we account for the fact that there is significant scatter
in the relation between central galaxy luminosity and halo mass, and 
that the satellite halo occupation function drops off more rapidly
than a power-law at low masses just above $M_\mathrm{min}$.
%
The satellite galaxy fraction is then
\begin{equation}
 f_\mathrm{sat} \,=\, \frac{
   \int_{M_\mathrm{min}} dM\,(dn/dM)\,\langle N_\mathrm{sat}|M,L_\mathrm{min} \rangle}{
   \int_{M_\mathrm{min}} dM\,(dn/dM)\,(\langle N_\mathrm{cen}|M\rangle +
     \langle N_\mathrm{sat}|M,L_\mathrm{min} \rangle)},
\end{equation}
where $dn/dM$ is the halo mass function.
The range of models that approximately fit the measurement (with $\chi_\mathrm{d.o.f.}^2<1$) 
have a satellite fraction of $f_\mathrm{sat}\approx32\%$.
This is fifty per cent larger than the satellite fraction of all galaxies with $M_r<-19.5$,
$f_\mathrm{sat}\approx24\%$.
%

There are two significant differences between the models that fit the correlation function
of red spirals in the figure and the fiducial model for all galaxies brighter than the luminosity threshold.
Firstly, and most importantly, the halo mass threshold $M_\mathrm{min}$ for red spirals is about 0.1-0.2 dex larger than that of the whole population, while the parameter $M_1$, which characterizes the amplitude of the mean halo occupation function and approximately corresponds to the mass of a halo that hosts one satellite on average, is only $\approx12$ times larger than $M_\mathrm{min}$, rather than the fiducial factor of $\approx20$.
The slope $\alpha$ of the halo occupation function is unchanged. 
If instead $\alpha$ were much larger for red spirals, indicating more galaxies in more massive haloes, and $M_1/M_\mathrm{min}$ were unchanged, then red spirals would be predominantly located in massive, cluster-sized haloes.  However, such models are ruled out by the measurement.
Since $M_1/M_\mathrm{min}$ has a lower value, the mean of the halo occupation distribution, 
$\langle N_\mathrm{sat}\rangle$, 
is relatively large even in lower mass haloes, 
indicating that the red spiral satellite galaxies
are expected to reside in haloes with a wide range in mass.
Although red spirals have been observed in the 
outskirts of galaxy clusters (hosted by high-mass haloes, with $M>10^{14}\,h^{-1}\,M_\odot$), 
we predict that they are located in galaxy groups as well (hosted by less massive haloes, with masses as low as $10^{13}\,h^{-1}\,M_\odot$).
Secondly, in order to fit the suppressed correlation function at scales of
$r_p<500\,\mathrm{kpc}/h$, models in which the satellite galaxy number density profile
is less concentrated (by a factor of three) than the dark matter profile are favored.
However, this parameter is not well-constrained.
%

We now address the possibility that our results are affected by dust reddening 
and galaxy inclinations.  
For example, a significant fraction of edge-on S0s could be mistakenly classified
as spirals, or star-forming edge-on spirals could be sufficiently dusty
to satisfy our colour criterion for `red spirals'.
Corrections for inclination-related effects are available in the literature
(Masters et al. 2003, Shao et al. 2007, Maller et al. 2008), and they are not negligible.
In Galaxy Zoo we can determine whether objects were classified as spiral due to 
the presence of visible spiral arms or simply due to a disky, edge-on appearance.
Following Bamford et al. (2009),
we define as `edge-on/unclear' those objects for which the majority of 
classifiers could not individually discern a spiral arm direction 
($p_\mathrm{CW}+p_\mathrm{ACW}<p_\mathrm{EU}$, where CW, ACW, and EU refer to 
clockwise, anti-clockwise, and edge-on/unclear, respectively).
Note that this is conservative, as even if a minority of classifiers can discern
spiral arms in a galaxy, then it is very likely to be a spiral.

\begin{figure}
 \includegraphics[width=\hsize]{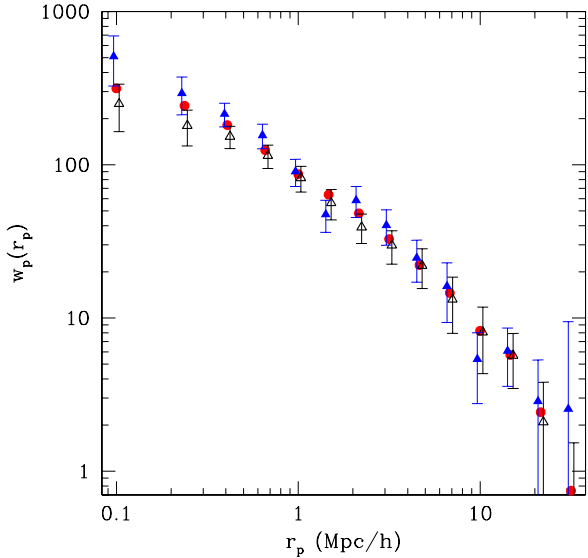} 
 \caption{Projected correlation function of red edge-on spirals (open triangles),
          red face-on spirals (filled triangles), and all red spiral galaxies 
          (red circles, same as Fig.~\ref{redspi}) in the $M_r<-19.5$ catalogue.
          The points have been slightly offset in $\mathrm{log}~r_p$, for clarity.}
 \label{incltest}
\end{figure}

In Figure~\ref{incltest}, we compare the clustering of red edge-on spirals 
(galaxies redder than the colour cut in Eqn.~\ref{colourcut}, with 
$p_\mathrm{CW}+p_\mathrm{ACW}<p_\mathrm{EU}$, and with $P_\mathrm{sp}>0.8$) and of 
red face-on spirals (redder than the colour cut, with $p_\mathrm{CW}+p_\mathrm{ACW}>p_\mathrm{EU}$,
and with $P_\mathrm{sp}>0.8$).
The red edge-on spirals outnumber the red face-on ones by a factor of more than 3:1.
The red edge-on spirals appear slightly less strongly clustered than the red face-on spirals,
consistent with the hypothesis that many of them are star-forming galaxies, which tend to be
located in underdense environments.  Nonetheless, the difference between the clustering 
measurements is of very weak statistical significance.
We have also used the apparent axis ratio ($b/a$) as an observational inclination indicator,
and have found that,
for red spiral galaxies,
it too does not depend strongly on the environment.  In addition,
the $b/a$ distributions of red spirals and blue spirals are similar, especially when the spiral arms are visible.
These tests show that galaxy inclinations can have only a minimal effect on our results,
and they do not affect our conclusions.
A detailed investigation of the effects of dust and inclination on spiral galaxies
in the Galaxy Zoo is currently underway.

\begin{figure}
 \includegraphics[width=\hsize]{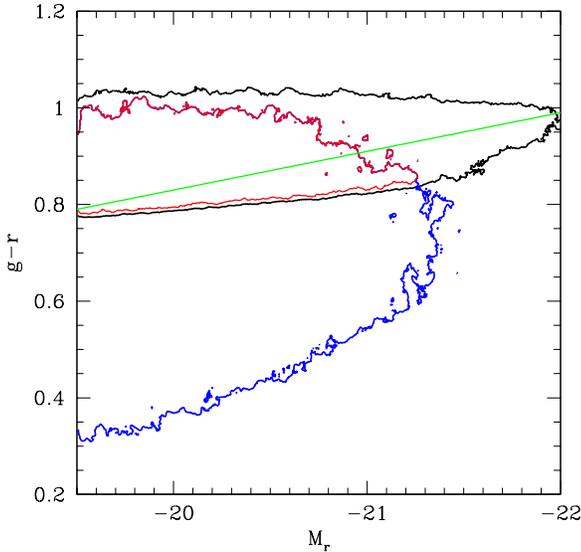} 
 \caption{$(g-r)\,-\,M_r$ colour-magnitude diagram for the $M_r<-19.5$ catalogue.
          Red contour circumscribes most of the red spiral galaxies;
          black and blue contours circumscribe most of the red galaxies and spiral galaxies,
          respectively.
          The contours encompass colour-magnitude points with galaxy counts that are
          5 per cent or more of the maximum count in the \textit{full} catalogue
          (which is peaked at $g-r=0.92$, $M_r=-19.9$), thus showing the relative
          contribution of each subset of the data.
          The satellite galaxy mean colour-magnitude sequence of Skibba \& Sheth (2009)
          is also shown (green line).
         }
 \label{redspiCMD}
\end{figure}

Finally, we show the colour-magnitude diagram of red spiral galaxies compared to that
of all red galaxies and all spiral galaxies in Figure~\ref{redspiCMD}.
Red spirals consist of $18$ per cent of the total population of red galaxies and also $23$ per cent of the spiral galaxy population.
Perhaps not surprisingly, red spirals are concentrated at the fainter, somewhat bluer
part of the red sequence.
The majority of spiral galaxies occupy the `blue cloud', although they have a wide
range in colour at fixed luminosity.

We interpreted the clustering result in Figure~\ref{redspi}, and in particular
the suppressed small-scale correlation function of red spiral galaxies, 
as evidence that red spirals are often satellite galaxies.
In the colour-magnitude diagram in Figure~\ref{redspiCMD}
we have also plotted the satellite galaxy mean colour-magnitude sequence of 
Skibba \& Sheth (2009) and Skibba (2009), which they found to be consistent 
with measurements of colour mark correlation functions in the SDSS, and with
the satellite colours as a function of luminosity and group richness in SDSS group catalogues.
Skibba (2009) showed that a large fraction of faint satellites are somewhat blue,
and argued that many are likely still undergoing processes of star formation quenching, while
the majority of brighter satellites tend to lie on the red sequence.

The fact that the satellite galaxy colour sequence in the figure goes through
the colour-magnitude region occupied by blue spirals at faint luminosities 
($M_r\geq-19.5$), red spirals at intermediate luminosities ($-21\leq M_r\leq-19.5$),
and red early-types at bright luminosities ($M_r<-21$), 
suggests that satellites tend to follow a particular evolutionary sequence.
Firstly, it suggests that the time-scale of the morphological transformation
of these galaxies is longer than that of the quenching of their star formation,
consistent with the analysis of Wolf et al. (2008). 
Secondly, it is useful to recall that while faint satellites can be found in dark matter haloes
of all masses, bright satellites tend to reside only in massive haloes (Skibba et al. 2007).
Therefore, while red spiral satellite galaxies may be potentially found in haloes of
a wide range in mass, red early-type satellites are rarer and tend to be found 
only in massive haloes, in the inner regions of galaxy clusters.
Consequently, we argue that many red spirals are satellites in groups and clusters,
in a variety of environments,
and were formally blue spirals whose star formation has been suppressed.
This picture is consistent with the analysis of van den Bosch et al. (2008a).

In contrast, central galaxies have a bluer mean colour sequence as a function of luminosity
than satellites.
At a given luminosity, central galaxies tend to be slightly bluer than satellites
and tend to be in less massive haloes, although central galaxies account for the
vast majority of luminous red galaxies, in more massive haloes (Skibba 2009).
We interpret the strong correlation between colour and environment for spiral galaxies
(Fig.~\ref{colourMCFellspi}) such that, most blue spirals are central galaxies
in low-mass haloes (\textit{i.e.}, in less dense environments), while red spirals
are either central galaxies in intermediate-mass haloes 
(with masses up to $M\sim10^{13}\,h^{-1}\,M_\odot$, corresponding to $M_r\sim-21$)
or satellites in more massive haloes (with masses $M\geq10^{13}\,h^{-1}\,M_\odot$, \textit{i.e.}, in denser environments).
The red spiral central galaxies tend to be located in galaxy groups, because of the
correlation between halo mass and group richness.
Following mergers with their own satellites, these central galaxies could
form bulges with supermassive black holes, and
transform their morphologies into lenticular and elliptical types (the GZ early-type class).
If these groups are accreted onto clusters, the former central galaxies would be the
origin of the red early-type satellites in massive haloes.
This picture is consistent with the analysis of Schiminovich et al. (2007), Bell (2008),
Pasquali et al. (2008), and Schawinski et al. (2009a).

The different relations between morphology and colour and between morphology and
group mass for central and satellite galaxies are currently being 
investigated further (Skibba et al., in prep.), with the galaxy group catalogue 
of Yang et al. (2007).


%
%
%
%

\section{Discussion}\label{discuss}

We have used volume-limited galaxy catalogues from the SDSS Data Release 6 
(Adelman-McCarthy et al. 2008) with data from the Galaxy Zoo (Lintott et al. 2008) 
to study the environmental dependence of galaxy morphology and colour.
The Galaxy Zoo project produced a large catalogue of visual morphological galaxy classifications,
yielding reliable morphological type likelihoods for hundreds of thousands of galaxies.
We measured projected two-point correlation functions and marked correlation
functions, and we used the latter to identify and quantify the environmental
correlations of morphological type likelihoods and $g-r$ colour on a wide
range of scales.

We summarize our results as follows:

\begin{itemize}

\item
We measured marked correlation functions using the marks $P_\mathrm{sp}$ and $P_\mathrm{el}$,
the likelihoods of a galaxy being a spiral and early-type, respectively.
These showed clear environmental trends, such that at all scales probed,
spiral galaxies tend to reside in less dense environments and early-type galaxies 
tend to be located in overdense environments.
Furthermore, the scale-dependence of the mark correlations is remarkably smooth,
indicating that the marks have very little noise --- that is, the Galaxy Zoo
morphological type likelihoods, which have been corrected for 
classification bias, are very precise.

\item
We also measured the $P_\mathrm{mg}$ (merger/interaction likelihood) 
marked correlation function, and found 
that it is strongly correlated with the environment at scales up to 
$r_p\approx170\,\mathrm{kpc}/h$.  That is, galaxies on these scales have a much
larger likelihood of showing evidence of mergers and interactions than average.
We compared this measurement to a specific star formation rate marked correlation
function measured with a catalogue with similar volume limits (Skibba, Wake \& Sheth, in prep.),
and showed that it too is strongly peaked at the same scale and is
uncorrelated with the environment at larger scales.
Naturally, merging and interacting galaxies are normally studied at very small scales,
at separations of $r_p<40\,\mathrm{kpc}/h$, which is closer than we can reliably probe.
However, our result shows either that the effects of galaxy interactions and bursts of
star formation persist out to separations of $r_p\approx170\,\mathrm{kpc}/h$ or more,
or that, for galaxies with neighbours on these scales, a significant fraction of their
neighbours are merging and star-bursting.
While we cannot conclusively claim that there is a causal connection between 
mergers and star formation, our result is consistent with the picture that
mergers and interactions between galaxies trigger star formation in them.

Studies that attempt to identify both galaxies in interacting systems
frequently miss interacting objects, due to selection criteria and
difficulties in determining line-of-sight separations, for example.
Measuring the mark correlation function, using the 
merging/interacting \textit{likelihood} 
as the mark, provides a complementary way of assessing the frequency and properties
of galaxy interactions, effective to larger separations, and including
all interacting objects, in a statistical sense.

\item
We have sought to disentangle the environmental dependence of morphology and colour
by measuring morphology marked correlation functions at fixed colour and
colour marked correlation functions at fixed morphology.
We found that blue galaxies show no correlation between morphology and environment
at all scales, and red galaxies are weakly correlated with the environment at 
small scales only ($r_p<500\,\mathrm{kpc}/h$).  The fact that this correlation 
occurs in the `one-halo term', at separations smaller than the size of massive 
dark matter haloes, is interpreted as an indication of morphology gradients within 
haloes: red galaxies in the outer regions of haloes tend to be spirals while those 
in the inner regions tend to be early-types.  Because of the correlation between 
galaxy colours and mass-to-light ratios (e.g., Baldry et al. 2006), this also
implies a morphology gradient for massive galaxies.

\item
In contrast, at fixed morphology, a relatively strong correlation between colour
and environment remains.  In particular, both spiral and early-type galaxies
show significant colour-environment correlations out to scales of $r_p\sim10\,\mathrm{Mpc}/h$,
and the colours of the latter have the stronger environmental dependence (see Figure~\ref{A4}).
That is, red spiral galaxies tend to be located in denser environments
than blue spirals, and red early-types tend to be located in the densest environments,
at any scale the environment is measured.
The fact that the morphology-environment correlation is very weak at fixed colour while
the colour-environment correlation remains strong at fixed morphology implies
that the morphology--density relation could be more accurately viewed as a
\textit{colour}-density relation, which is consistent with the conclusions of
other studies (Blanton et al. 2005, Wolf et al. 2007, Bamford et al. 2009).

\item
We used the bimodal $g-r$ colour distribution as a function of $r$-band luminosity
(Skibba \& Sheth 2009) to estimate colour likelihoods of galaxies, $P_\mathrm{red}$
and $P_\mathrm{blu}$, analogous to the morphology likelihoods $P_\mathrm{el}$
and $P_\mathrm{sp}$.  We used these four likelihoods together as marks,
and found that the marked correlation function with the $P_\mathrm{sp} P_\mathrm{red}$
mark is significantly above unity.  That is, the opposite environmental correlations
did not cancel out; instead, the environmental dependence of colour is much stronger
than that of morphology.

\item
This result shows that galaxies that are both red and spiral tend to be located in 
dense environments.  These red spiral galaxies are not insignificant: they
are $9$ per cent of the galaxies in the $M_r<-19.5$ volume-limited catalogue, and they
account for nearly a quarter of all the spirals.  We compared the correlation
function of red spiral galaxies to that of all red galaxies and of spirals, and 
found that the clustering of red spirals is similar to that of red galaxies
on large scales, but is depressed on scales of $r_p<2\,\mathrm{Mpc}/h$, in the 
one-halo term, indicating that they have a high satellite fraction.
This is more evidence that many red spiral galaxies are satellite galaxies.
Skibba \& Sheth (2009) and Skibba (2009) showed that satellite galaxies
tend to follow a particular sequence on the colour-magnitude diagram, and it
passes through the region of the diagram occupied by red spirals.
These results suggest that satellite galaxies evolve a particular way:
faint satellites are accreted as blue late-types, and then they experience 
`strangulation', in which their hot gas reservoirs are stripped, causing 
their star formation to be gradually suppressed 
while they retain spiral morphology.
%

\end{itemize}


\section*{Acknowledgements}
We thank Aday Robaina, Anna Gallazzi, Frank van den Bosch, Ravi Sheth, and David Hogg for valuable discussions about our results. 
We also thank the anonymous referee for a careful reading of the paper and 
for helpful suggestions that improved it.

We thank Jeffrey Gardner, Andrew Connolly, and Cameron McBride
for assistance with their $N$tropy code, which was used 
to measure all of the correlation functions presented here.
$N$tropy was funded by the NASA Advanced Information Systems 
Research Program grant NNG05GA60G.

This work has depended on the participation of many members of the public in
visually classifying SDSS galaxies on the Galaxy Zoo website.  We thank them
for their extraordinary efforts in making this project a success.  We are
also indebted to various members of the media, both traditional and online,
for helping to bring this project to the public's attention.

Funding for the SDSS and SDSS-II has been provided by the 
Alfred P. Sloan Foundation, the Participating Institutions, 
the National Science Foundation, the U.S. Department of Energy, 
the National Aeronautics and Space Administration, 
the Japanese Monbukagakusho, the Max Planck Society, 
and the Higher Education Funding Council for England. 
The SDSS Web Site is http://www.sdss.org/.

The SDSS is managed by the Astrophysical Research Consortium for 
the Participating Institutions. The Participating Institutions are 
the American Museum of Natural History, Astrophysical Institute 
Potsdam, University of Basel, Cambridge University, 
Case Western Reserve University, University of Chicago, 
Drexel University, Fermilab, the Institute for Advanced Study, 
the Japan Participation Group, Johns Hopkins University, 
the Joint Institute for Nuclear Astrophysics, 
the Kavli Institute for Particle Astrophysics and Cosmology, 
the Korean Scientist Group, the Chinese Academy of Sciences (LAMOST), 
Los Alamos National Laboratory, 
the Max-Planck-Institute for Astronomy (MPA), 
the Max-Planck-Institute for Astrophysics (MPIA), 
New Mexico State University, Ohio State University, 
University of Pittsburgh, University of Portsmouth, 
Princeton University, the United States Naval Observatory, 
and the University of Washington.

\appendix

\renewcommand{\thefigure}{\Alph{appfig}\arabic{figure}}
\setcounter{appfig}{1}

\section{Mark Distributions and Their Effect on the Marked Correlation Functions}\label{markdistapp}

Mark correlation functions are sensitive to the distributions of the marks.
In this appendix, we show the mark distributions for the measurements in
Section~\ref{morphcolour}, in which we compared the environmental dependence
of galaxy morphology and colour, and we show how the mark correlation measurements
are affected when one accounts for the different mark distributions.
 
\begin{figure}
 \includegraphics[width=\hsize]{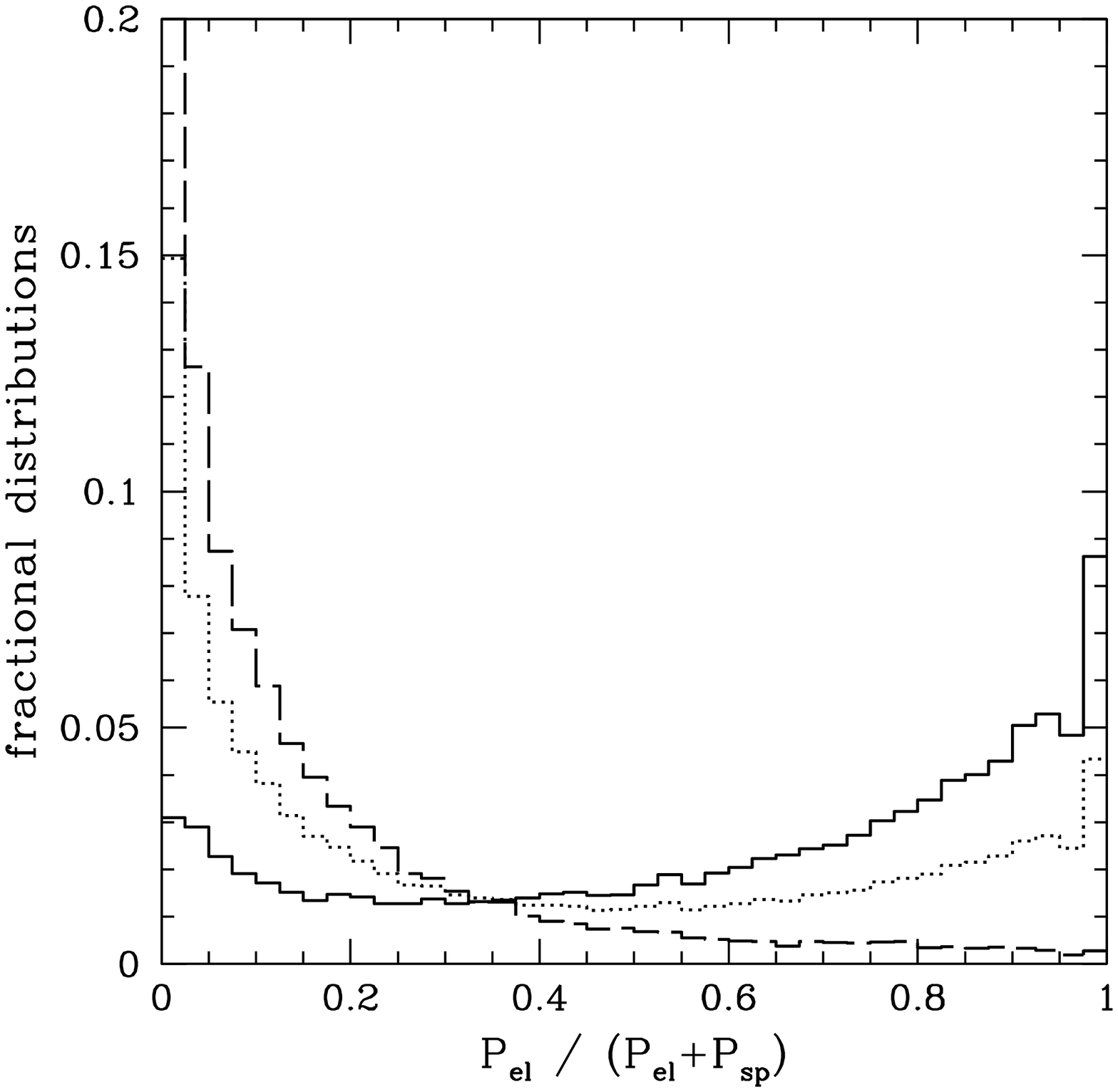} 
 \caption{$P_\mathrm{el}/(P_\mathrm{el}+P_\mathrm{sp})$ distributions of
          all galaxies (black dotted histogram), red galaxies (red solid histogram),
          and blue galaxies (blue dashed histogram), 
          with $P_\mathrm{red}>0.8$ and $P_\mathrm{blu}>0.8$,
          in the $M_r<-19.5$ volume-limited catalogue.}
 \label{A1}
\end{figure}

In Figure~\ref{A1}, we show the distribution of $P_\mathrm{el}$
morphology likelihood (normalized by $P_\mathrm{el}+P_\mathrm{sp}$),
for red and blue galaxies, where `red' and `blue'
refers to the colour likelihoods estimated from the $g-r$ colour distribution
as a function of $r$-band luminosity (Skibba \& Sheth 2009), for $P_\mathrm{red}>0.8$
and $P_\mathrm{blu}>0.8$.
Not surprisingly, a larger fraction of red galaxies than blue galaxies are early-types,
and a larger fraction of blue galaxies are spirals.

In order to account for the different mark distributions for the mark correlation
functions of red and blue galaxies, we rescale their $P_\mathrm{el}$
mark distributions so that they have the same distribution as that of all
galaxies (black histogram in Fig.~\ref{A1}).
The procedure is simple: if there are $N$ marks in a catalogue, then we 
generate $N$ random marks such that they have the required distribution.
Then we sort these marks and replace the original marks in the catalogue with them.
We maintain the rank order of the galaxies' marks; only their distribution has changed.

\begin{figure}
 \includegraphics[width=\hsize]{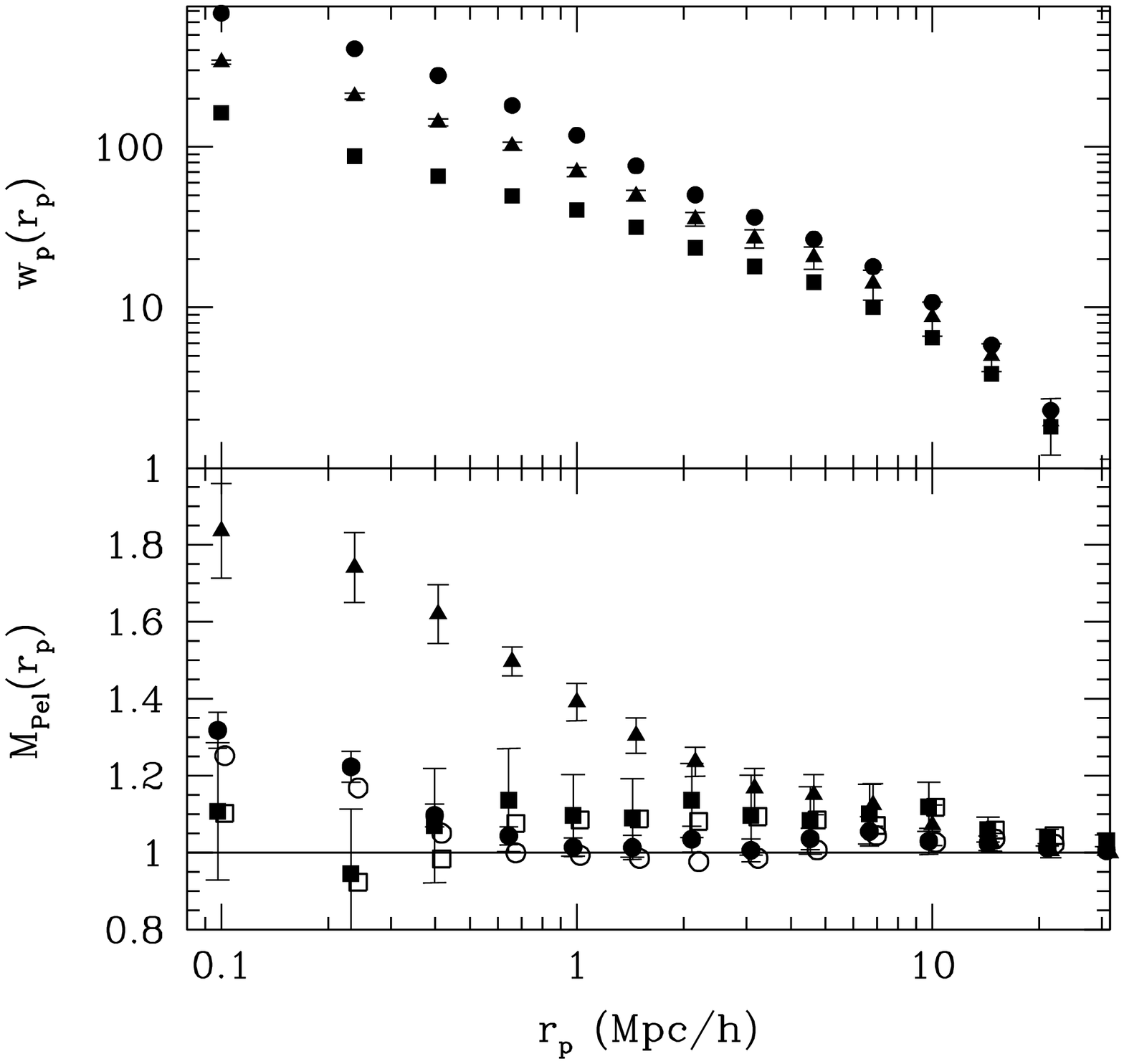} 
 \caption{$P_\mathrm{el}$ marked correlation function (lower panel) of all
          galaxies in the $M_r<-19.5$ catalogue (black triangles) compared
          to that of red and blue galaxies (red circle points and blue square points), as in Figure~\ref{morphMCFredblu}, but with the mark
          distributions forced to match the black dotted histogram in Figure~\ref{A1}.
          `Red' and `blue' galaxy subcatalogues are determined using the colour-magnitude cut in
          Eqn~\ref{colourcut} (filled points) or using the colour likelihood
          thresholds $P_\mathrm{red}>0.8$ and $P_\mathrm{blu}>0.8$ (open points).
          The positions of the galaxies are unchanged, so the unmarked correlation
          functions (upper panel) are the same as before.}
 \label{A2}
\end{figure}

Following this procedure, we repeated the measurements in Figure~\ref{morphMCFredblu}
with the rescaled marks for red and blue galaxies.
The result is shown in Figure~\ref{A2}, and it is almost exactly the same.
The morphology gradient of red galaxies on small scales is slightly stronger.
There is a hint of a positive environmental correlation for blue galaxies,
which could be interpreted as an indication that blue early-types tend to
reside in denser environments than blue spirals, but this trend is
of weak significance (only $\sim1-\sigma$).

\begin{figure}
 \includegraphics[width=\hsize]{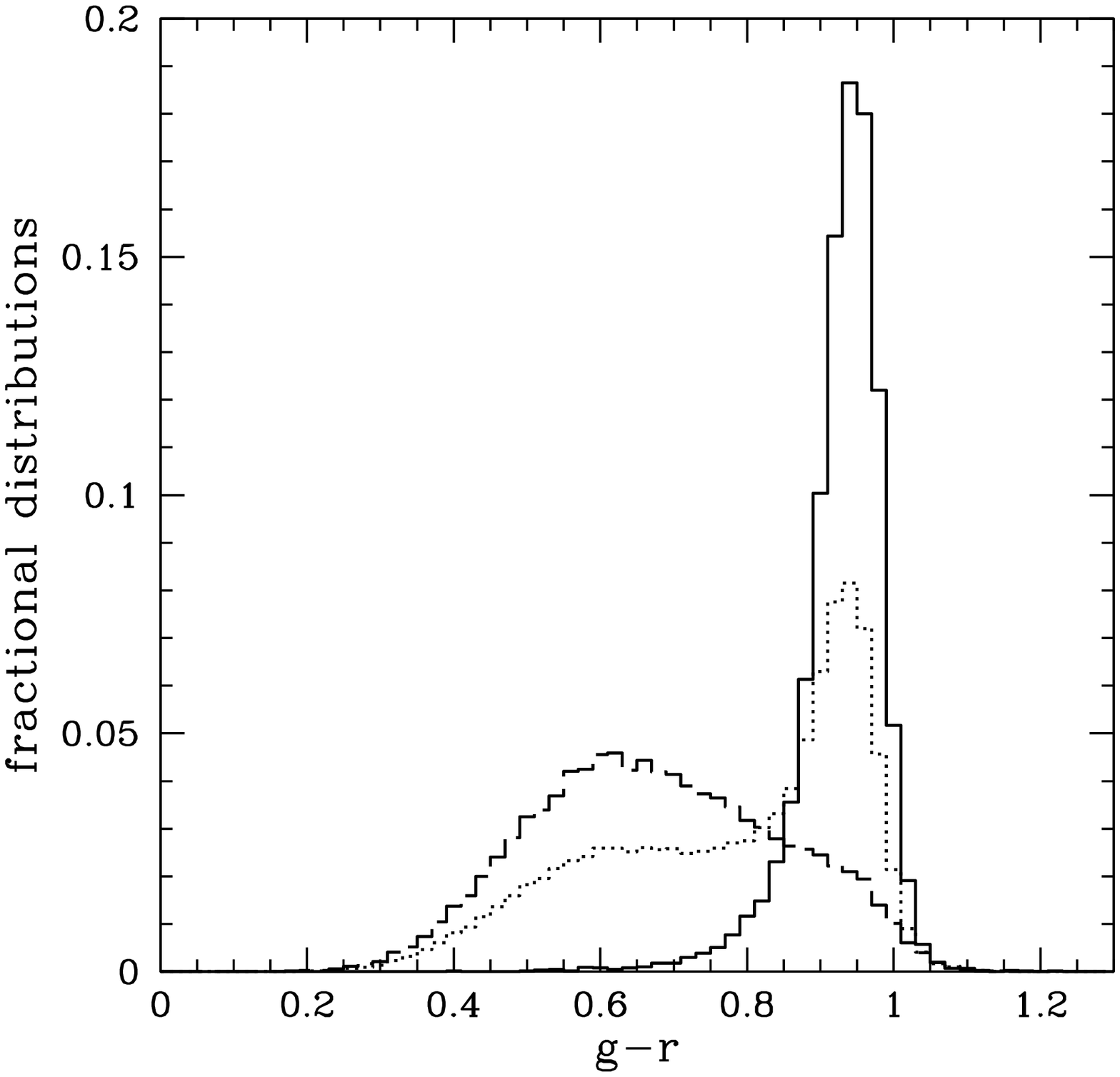} 
 \caption{$g-r$ colour distributions of all galaxies (black dotted histogram),
          early-type galaxies (red solid histogram), and spiral galaxies 
          (blue dashed histogram), in the $M_r<-19.5$ volume-limited catalogue.
          Spiral and early-type galaxies are defined with $P_\mathrm{sp}>0.8$
          and $P_\mathrm{el}>0.8$, respectively.}
 \label{A3}
\end{figure}

Next, we show the distribution of $g-r$ colour for spiral and early-type galaxies,
with $P_\mathrm{sp}>0.8$ and $P_\mathrm{el}>0.8$, in Figure~\ref{A3}.
The distribution for early-types is strongly peaked on the red sequence, 
while spirals have a wide and smooth distribution, with a significant red fraction.

\begin{figure}
 \includegraphics[width=\hsize]{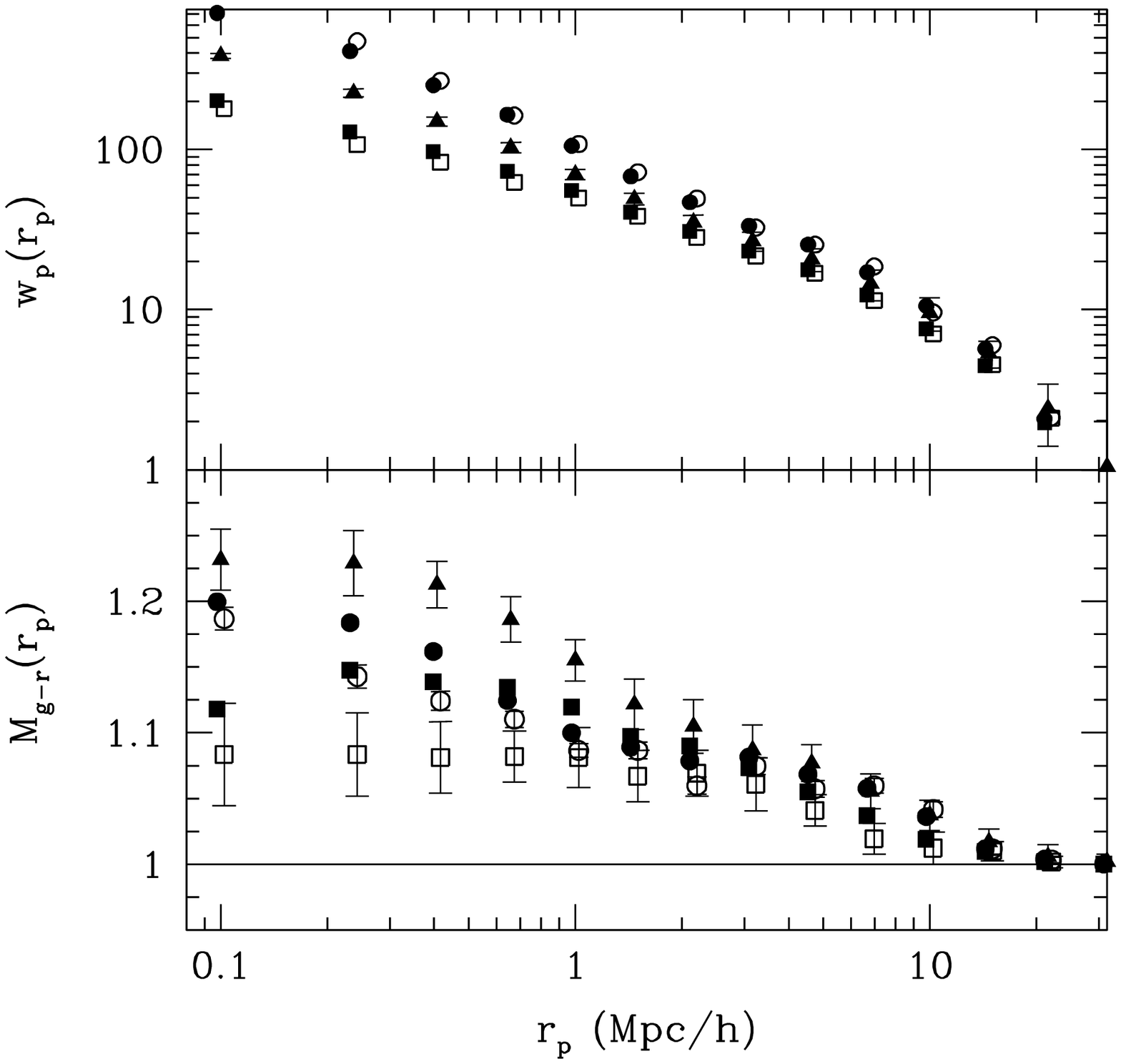} 
 \caption{$g-r$ colour marked correlation function (lower panel) of all
          galaxies in the $M_r<-19.5$ catalogue (triangle points) compared to
          that of spiral and early-type galaxies (red circle points and blue square points), as in Figure~\ref{colourMCFellspi}, but with the
          mark distributions forced to match the black dotted histogram in Figure~\ref{A3}.
          Spiral and early-type subcatalogues are determined using the greater morphology
          likelihood as the distinguishing criterion (filled points)
          or using the likelihood threshold of $P>0.8$ (open points). }
 \label{A4}
\end{figure}

We now rescale the colour mark distributions, with the same procedure as before,
forcing the spiral and early-type subsamples to have the same colour distribution
as that of all galaxies (black dotted histogram in Fig.~\ref{A3}).
The resulting colour marked correlation functions are shown in Figure~\ref{A4}.
This result clearly shows that red early-type galaxies are more strongly
clustered than red spirals, especially within the one-halo term.
The apparent opposite trend in Figure~\ref{colourMCFellspi}
was simply due to the extremely narrow colour distribution of early-types
(\textit{i.e.}, the colours were not allowed to depart significantly from the mean,
so the colour marks of strongly clustered red early-types were not much larger than
the colour marks of bluer early-types in less dense environments).

\begin{figure}
 \includegraphics[width=\hsize]{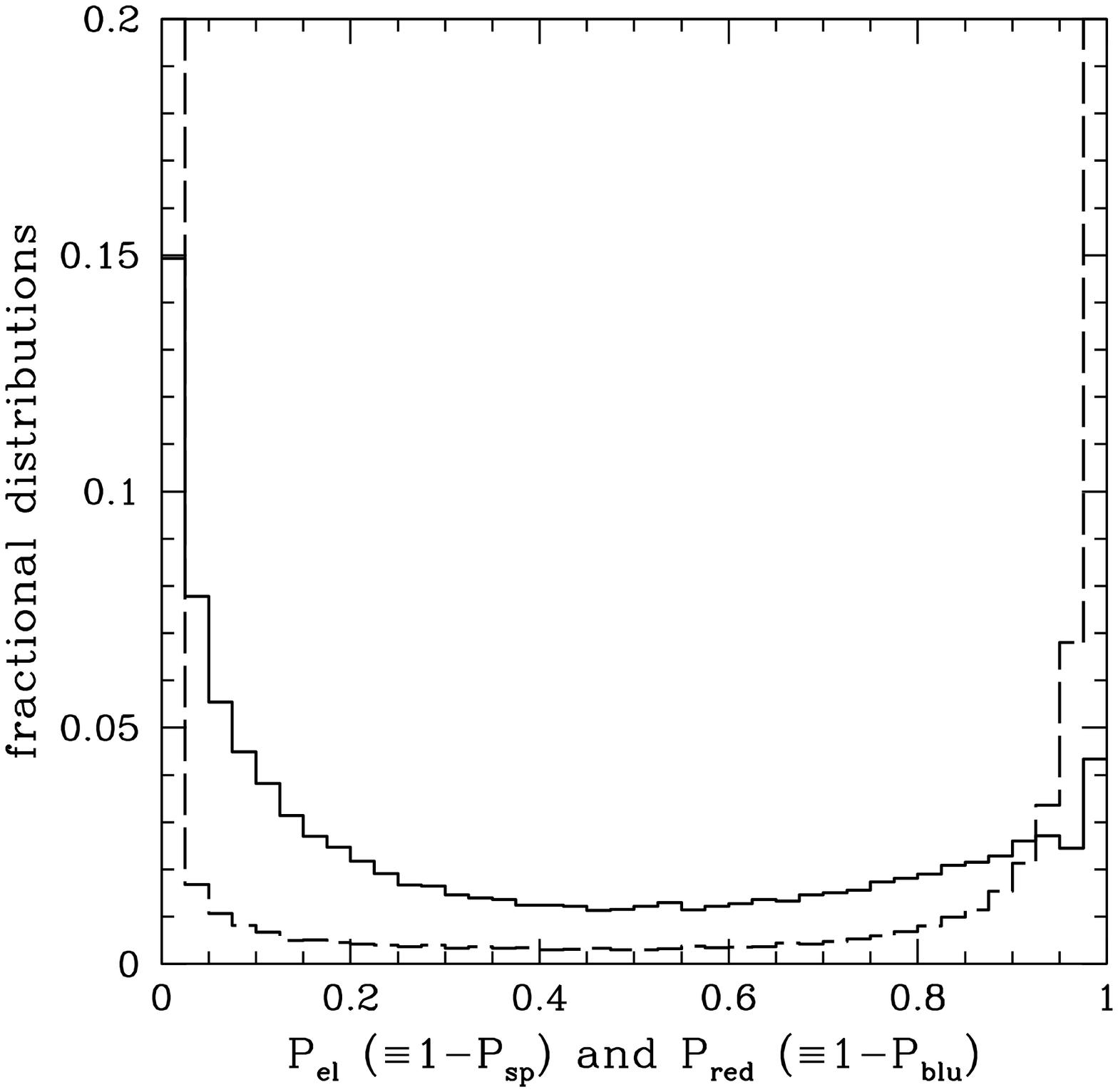} 
 \caption{Distribution of $P_\mathrm{el}$ (solid black histogram), 
          normalized so that $P_\mathrm{sp}+P_\mathrm{el}=1$,
          and $P_\mathrm{red}$ (red dashed histogram), normalized so that
          $P_\mathrm{blu}+P_\mathrm{red}=1$, in the $M_r<-19.5$ volume-limited catalogue.}
 \label{A5}
\end{figure}

Finally, we compare the distribution of $P_\mathrm{el}$,
normalized by $P_\mathrm{el}+P_\mathrm{sp}$, to that of
$P_\mathrm{red}$ in Figure~\ref{A5}.
The $P_\mathrm{red}$ distribution is more strongly peaked near 0 and 1 than that of 
$P_\mathrm{el}$, simply because the colour distribution of galaxies is strongly 
bimodal, with an underpopulated `green valley', across much of the range of 
luminosities (Skibba \& Sheth 2009).

\begin{figure}
 \includegraphics[width=\hsize]{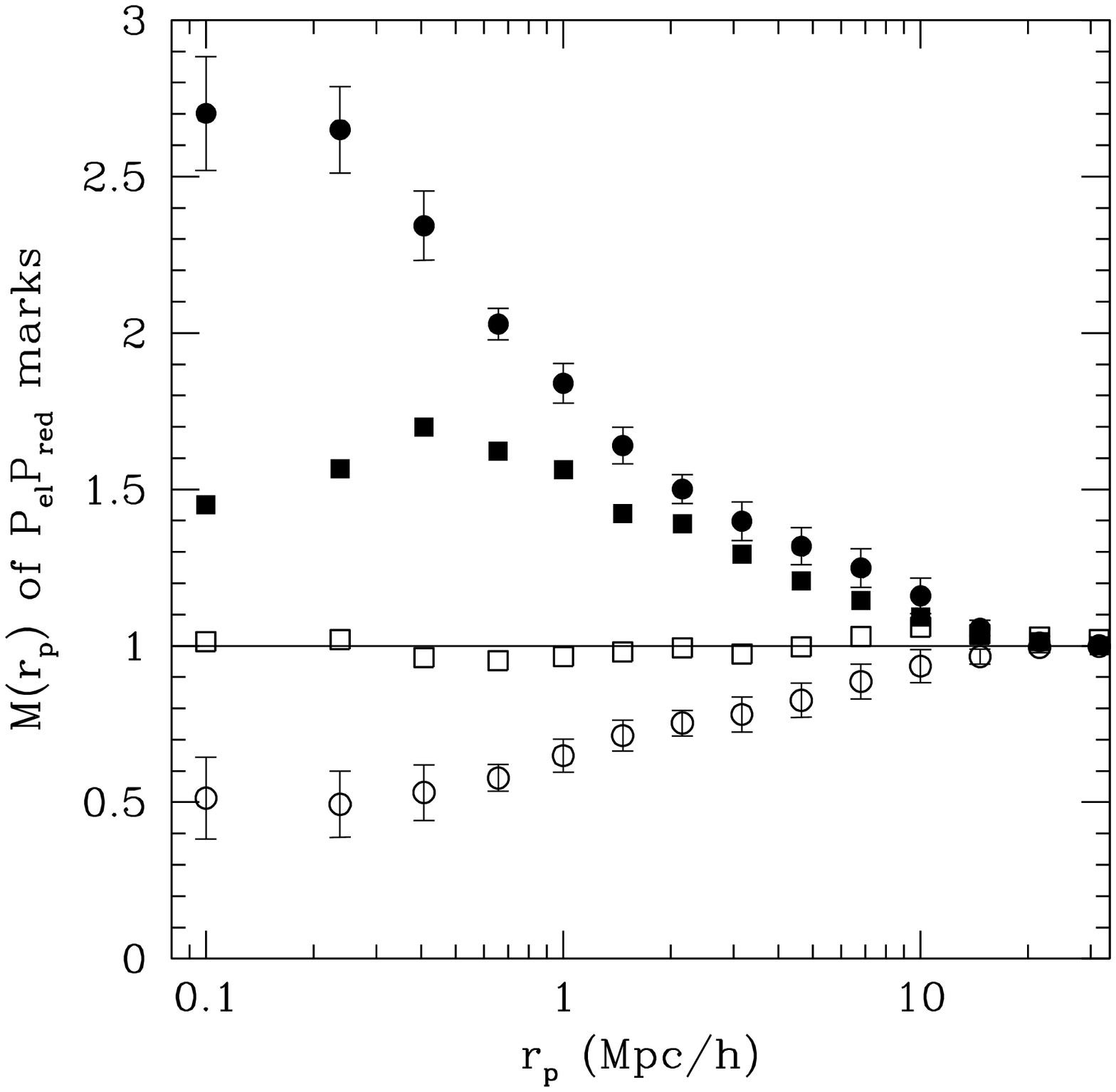} 
 \caption{Marked correlation functions for the four colour-morphology combinations,
          $P_\mathrm{el}P_\mathrm{red}$ (red filled circles), $P_\mathrm{sp}P_\mathrm{red}$
          (red filled squares), $P_\mathrm{el}P_\mathrm{blu}$ (blue open squares),
          and $P_\mathrm{sp}P_\mathrm{blu}$ (blue open circles).
          as in Figure~\ref{PredPspiusw}, but the distribution of $P_\mathrm{red}$
          (and $1-P_\mathrm{blu}$) is forced to match that of $P_\mathrm{el}$
          in Figure~\ref{A5}.}
 \label{A6}
\end{figure}

Since the distribution of colour likelihoods is more strongly peaked at the minimum and maximum
than the distribution of morphology likelihoods, it is possible that
this could artificially strengthen the environmental dependence of colour 
vis-\'{a}-vis that of morphology.
We account for this by maintaining the rank order of $P_\mathrm{red}$ of the galaxies
in the catalogue, and giving them the same distribution as that of $P_\mathrm{el}$.
We then re-measured the marked correlation functions with the four colour-morphology
combinations of $P_\mathrm{el}$, $P_\mathrm{sp}$, $P_\mathrm{red}$, and $P_\mathrm{blu}$
shown in Figure~\ref{PredPspiusw}.
Figure~\ref{A6} shows the results, and they are almost the same.
The only significant difference is that the $P_\mathrm{blu} P_\mathrm{el}$
mark appears to be uncorrelated with the environment at all scales,
possibly due to the fact that blue early-types are extremely rare ($0.7$ per cent).
However, our conclusion that galaxies with high $P_\mathrm{red}$ and $P_\mathrm{sp}$
tend to reside in denser environments than average is robust.

\label{lastpage}

\end{document}